\documentclass[12pt]{article}
\usepackage[utf8]{inputenc}

\usepackage{amsmath,amssymb}
\usepackage{graphicx}
\usepackage{nicefrac}
\usepackage{comment}

\usepackage{hyperref}
\usepackage{url}
\usepackage[dvipsnames]{xcolor}
\usepackage{rotating}
\usepackage[super]{nth}
\usepackage{graphicx}
\usepackage{authblk}
\usepackage{pdflscape}

\usepackage[round,sort,authoryear]{natbib}
\usepackage[left=1.0in, right=1.0in, top=1in, bottom=1in]{geometry}

\usepackage{enumerate}

\usepackage{newtxtext,newtxmath}
\usepackage{caption}
\usepackage{subcaption}

\newcommand\mycaption[2]{\caption{\textbf{#1}\newline\small#2}}
\newcommand{\sym}[1]{{#1}} 

\graphicspath{{figures/}} 
\title{Evaluating the principle of relatedness: Estimation, drivers and implications for policy}

\author[1,*]{Yang Li} 
\author[2,1,**]{Frank Neffke}   
\affil[1]{\small Center for International Development, Harvard University - 79 JFK street, 02138 Cambridge MA, USA}
\affil[2]{\small Complexity Science Hub Vienna - Josefst{\"a}dter Stra\ss e 39, 1080 Vienna, Austria}
\affil[*]{\texttt{Yang\_Li@hks.harvard.edu}}
\affil[**]{\texttt{Frank\_Neffke@hks.harvard.edu}}

\date{\today}

\begin{document}

\maketitle

\begin{abstract}
    A growing body of research documents that the size and growth of an industry in a place depends on how much related activity is found there. This fact is commonly referred to as the ``principle of relatedness''. However, there is no consensus on why we observe the principle of relatedness, how best to determine which industries are related or how this empirical regularity can help inform local industrial policy. We perform a structured search over tens of thousands of specifications to identify robust  -- in terms of out-of-sample predictions -- ways to determine how well industries fit the local economies of US cities. To do so, we use data that allow us to derive relatedness from observing which industries co-occur in the portfolios of establishments, firms, cities and countries. Different portfolios yield different relatedness matrices, each of which help predict the size and growth of local industries. However, our specification search not only identifies ways to improve the performance of such predictions, but also reveals new facts about the principle of relatedness and important trade-offs between predictive performance and interpretability of relatedness patterns. We use these insights to deepen our theoretical understanding of what underlies path-dependent development in cities and expand existing policy frameworks that rely on inter-industry relatedness analysis.
\end{abstract}

\subsection*{Funding}
Frank Neffke acknowledges financial support from the Austrian Research Agency [FFG, project \#873927:
ESSENCSE].

\newpage
\section{Introduction}

The field of human geography has uncovered a number of striking empirical regularities, such as Zipf's law for city-size distributions \citep{auerbach1913gesetz,zipf1946p}, the law of gravity for social interactions \citep{tinbergen1962shaping}, Tobler's first law of geography for spatial dependence \citep{tobler1970computer} and the urban wage and productivity premiums for the exceptional role that cities play in the economy \citep{rosenthal2004evidence,bettencourt2007growth}. Recently, a new regularity has been proposed: the \emph{principle of relatedness} \citep{hidalgo2018principle}. According to this principle, the rate of growth of an activity in a location can be predicted from the prevalence of related activities in that same location. The  principle of relatedness has not only received ample attention in scholarly work, but also plays an important role in economic policy frameworks that provide the basis for large-scale place-based development programs \citep{balland2019smart,rigby2022eu,boschma2022evolutionary,hidalgo2022policy}, such as the EU's smart specialization policy \citep{foray2009smart}. However, important aspects of the principle of relatedness remain poorly understood. First, when researchers examine the principle of relatedness, they face a large number of \emph{ad hoc} choices for how to construct and then use relatedness measures. Second, we know little about \emph{when} and \emph{why} we observe the principle of relatedness. Third, most existing policy frameworks that leverage the principle of relatedness do not incorporate this methodological and conceptual uncertainty in their policy recommendations. In this paper, we perform a structured search over tens of thousands of specifications of models that aim at quantifying the strength of the principle of relatedness. Doing so, we (1) provide practical guidance for empirical research, (2) uncover a number of substantive insights into the principle of relatedness and the forces giving rise to it; and (3) complement existing policy frameworks by highlighting existing blind spots and proposing an alternative way to use the principle of relatedness in policy that focuses on the identification, not of growth opportunities, but of developmental bottlenecks.

The principle of relatedness traces its intellectual roots to debates on how local economies reinvent themselves \citep[e.g.,][]{jacobs1969economy,grabher1993weakness,glaeser2005reinventing,martin2006path}. For cities to stay relevant in a world where technologies and competitive forces constantly change, their economies need to find new growth paths. In this context, \cite{jacobs1969economy} differentiated between economic growth and economic development. The former refers to increases in efficiency: as cities become better at utilizing their existing resources, their productivity grows. The latter refers to economic renewal and diversification. According to \citeauthor{jacobs1969economy}, cities that only raise their efficiency risk deep crises when technological paradigms change or competitive forces shift. Canonical examples can be found in the developmental histories of Detroit in the U.S., Manchester in the U.K. and the Ruhr area in Germany. The lack of renewal in such regions would later give rise to an extensive literature on path dependence and regional lock-in \citep[e.g.][]{grabher1993weakness,martin2006path}. 

An important conclusion of this research is that, to avoid decline, the successful regions of past epochs need to diversify into new activities. However, such new growth paths do not arise out of thin air. Instead, as \citeauthor{jacobs1969economy} put it, cities grow by ``adding new work to old'' \citep{jacobs1969economy}: new economic activities are often related to what a city already knows how to do. 

Although the idea that local economies develop by branching into activities related to their current strengths \citep[e.g.][]{frenken2007theoretical} has immediate intuitive appeal, empirically validating this hypothesis initially ran into a serious obstacle: how do we decide which activities are related? It would take several decades before this issue had been resolved and Jacobs' claims backed by quantitative evidence. 

A breakthrough emerged with the insight that important information can be gained from studying the portfolios of activities in which economic entities choose to be active. Accordingly, certain combinations of economic activities give rise to \emph{economies of scope}: there are advantages to engage in them simultaneously. When myriad economic agents make micro-level portfolio decisions aimed at exploiting these economies of scope, such advantages leave a trace in the activity mix of economic entities. In other words, relatedness reveals itself in the tendency of activities to co-occur in productive portfolios.\footnote{This insight was first leveraged by scholars in scientometrics \citep{engelsman_mapping_1991} and strategic management \citep{teece1994understanding}.} Exploiting this insight allowed accumulating evidence  in support of Jacobs' claims across a wide variety of datasets and contexts, successfully connecting the growth of an industry \citep{porter2003economic,neffke2011regions,essletzbichler2012evolutionary,florida2012geographies,delgado2014clusters}, export category  \citep{zhu2017jump}, occupation \citep{muneepeerakul2013urban}, technology \citep{boschma2015geography,petralia2017climbing} or academic field \citep{guevara2016research} to the local prevalence of related activities.\footnote{For a recent overview, see \cite{hidalgo2018principle}.} 

These empirical studies typically proceed in three steps. First, they determine the relatedness among economic activities. Next, for each activity in a region, they calculate how prevalent related activities are in that region. Finally, they regress the growth rates of an activity on the prevalence of related activities.

Although this procedure seems straightforward, its implementation involves several \emph{ad hoc} choices. In this paper, we undertake a structured exploration of tens of thousands of candidate specifications that aim to replicate the principle of relatedness. To do so, we create a specification grid on which we vary several aspects of the empirical model design. First, economic activities can be observed in different types of economic entities, each of which can be used to count co-occurrences. For instance, \cite{hidalgo2007product} study the co-occurrence of traded product categories in the export mix of countries, \cite{porter2003economic} and \cite{delgado2010clusters} of industries in regions, \cite{teece1994understanding} and \cite{bryce2009general} of industries in firms, and \cite{neffke2011regions} of products in manufacturing plants. Second, information on productive portfolios can be turned into relatedness matrices in different ways. Third, given a relatedness matrix, there are many ways to quantify how related one activity is to the local economy as a whole. For each of these design aspects, we explore dozens of choices, yielding tens of thousands of candidate specifications. 

Our empirical analysis relies on Dun and Bradstreet's World Base (henceforth, ``D\&B data''). This dataset reports for over 100 million establishments worldwide the number of employees, geographical coordinates, headquarter-subsidiary relations and up to six economic activities. We choose this dataset, because it allows us to observe co-occurrences of economic activities at four different levels of aggregation: the establishment, the firm, the city and the country. To test which candidate specification in our specification grid best captures the principle of relatedness, we focus on the economies of US cities and aggregate the D\&B data to the level of city-industry combinations. Next, we repeatedly divide the data into train and test samples and generate out-of-sample predictions for city-industry employment and employment growth patterns. This allows us to rank all candidate specifications by their out-of-sample predictive performance. 

Overall, we find broad support for the principle of relatedness: many specifications corroborate the positive association between the presence of related economic activity and an industry's local size and growth rate. However, many specifications perform poorly, with about 10\% - 20\% of specifications failing to outperform benchmarks that use random relatedness matrices. Moreover, which specification to use depends to some extent on the productive unit in which we observe industry co-occurrences and different units give rise to different relatedness matrices. To support future research in this area, we provide a number of guidelines in terms of specification choices that should be avoided, as well as our preferred specifications. 

Apart from these specification recommendations, our analysis reveals a number of stylized facts about the principle relatedness. First, the principle of relatedness is often regarded as an expression of which capabilities a city has. Accordingly, moving into related activities is \emph{cheaper}, because it limits the number of new capabilities that the city needs to  acquire \citep[see, e.g., ][]{balland2019smart}. Therefore, many authors exclude from their analysis industries whose location decisions are not primarily driven by access to local capabilities, such as extractive industries, industries in the public sector and non-traded services. However, the principle of relatedness also turns out to be predictive in these sectors. This suggests that the principle of relatedness exists for reasons that go beyond the common explanation that inter-industry relatedness reflects commonalities in capability requirements. Second, we find that constructing relatedness matrices in different types of productive units reveal different types of economies of scope. For instance, labor-sharing rationales are an important factor explaining co-occurrences in establishments and firms, but much less so in cities or countries. Yet, the co-occurrence patterns in cities contain information for predicting location and growth patterns beyond what is expressed in establishment and firm level co-occurrences. Finally, good predictive performance does not not guarantee a clear understanding of why some local industries thrive in a city: some of the best performing specifications yield cluttered inter-industry relatedness networks that make it hard to disentangle clusters of related industries. 

Finally, when it comes to policy implications, an important finding is that the principle of relatedness is much better at predicting where industries are located than where they grow. As a result, if we were to prioritize industries by their predicted growth potential, we would often pick industries that won't exhibit growth spurts and miss industries that will. Moreover, even if we could identify promising growth candidates, the principle of relatedness offers little guidance for how to promote their growth. In fact, one may ask: if these industries fit the local economy so well, why haven't they grown yet? We will argue that this type of question can complement current policy frameworks by using the principle of relatedness, not to prioritize industries, but as an \emph{anomaly detection tool}. Accordingly, the principle of relatedness can help policy makers diagnose their economy, by prompting them to ask why specific industries are surprisingly small or large in their city. We conclude our paper by proposing to use this type of analysis to identify binding constraints to economic development in a city, integrating the principle of relatedness into the wider policy framework of \emph{Growth Diagnostics} \citep{hausmann2008growth}.

\section{Data}\label{sec:data}
The D\&B data are provided by Dun and Bradstreet, a business analytics firm. They contain information on over 100 million establishments across the world and offer an almost complete census of economic establishments in the U.S.. For each establishment, the dataset records an identifier (the so-called D.U.N.S.\footnote{D.U.N.S is a recursive acronym for D.U.N.S. Universal Numbering System.} number). This D.U.N.S. number is unique to the establishment and remains unchanged throughout its existence, regardless of changes in ownership. Furthermore, the dataset offers for each establishment the number of employees,\footnote{Outside the U.S., most employment figures are based on estimates by D\&B.} geographical coordinates, and the D.U.N.S. number of the parent establishment if the establishment is part of a larger corporation. The latter allows us to reconstruct corporate hierarchies that express these ownership linkages. Finally, each establishment can list up to six different industries. These industries are ordered by their importance, with the primary codes identifying the establishment's main industry. Industry codes are recorded in the SIC or NAICS classifications. In this paper, we rely on the 2011 and 2019 waves of the D\&B data, the first and last waves available to us at the time of the analysis. Because our 2011 wave only contains SIC codes, our analysis is based on 3-digit industries of this classification system.\footnote{At this level, there are over 400 industry codes, distinguishing, for instance, between the construction industries of ``Masonry, stonework, tile setting, and plastering'' and ``Plumbing, heating and air-conditioning'' or the manufacturing industries of ``Computer and office equipment'' and ``Household appliances''.} Table \ref{tab:summarystat} provides some general statistics for the dataset.

\begin{table}[t!]
    \begin{center}
        \begin{tabular}{lrrr}
\hline 
                                          variable &      counts &   mean &     std \\
\hline 
                                       \# employees &   165,987,254 &        &         \\
                                   \# establishments &    26,374,079 &        &         \\
         \# establishment ($>$1 SIC 3-digit industry) &     1,688,984 &        &         \\
                                           \# firms &      363,620 &        &         \\
          \# firm ($>$1 primary SIC 3-digit industry) &       83,548 &        &         \\
                                         \# cities &        927 &        &         \\
                                       \# countries &         100 &        &         \\
                                      \# industries &         415 &        &         \\
                    \# industries per establishment &             &   1.08 &    0.34 \\
\# industries per multi-industry establishment &             &   2.25 &    0.59 \\
                             \# industries per firm &             &   1.66 &    2.20 \\
\# industries per multi-industry firm  &             &   3.14 &    3.42 \\
                           \# industries per city &             & 288.37 &   56.11 \\
                          \# industries per country &             & 316.11 &   78.72 \\
                   \# employees per city-industry &             & 620.93 & 4,714.69 \\
              log(\# employees per city-industry) &             &   3.94 &    2.14 \\
              $\Delta \log( E_{ir})$  &             &  -0.08 &    0.79 \\
\hline 
\end{tabular}

    \end{center}
  \mycaption{Summary statistics}{Multi-industry establishments are establishments with at least two distinct 3-digit SIC codes. Multi-industry firms are firms that list establishments with at least two distinct primary 3-digit SIC codes. $\Delta \log( E_{ir})$  refers to employment growth between 2011 and 2019.}\label{tab:summarystat}
\end{table}

The D\&B data are highly representative of the US economy (a comparison with US County Business Patterns data is listed in Appendix \ref{sub:us_compare}), but not necessarily of other economies. Therefore, we will mainly work with US data, limiting the sample to US establishments and defining firms as sets of US establishments that report to the same domestic (US-based) parent. However, to examine the industrial portfolios of countries, we aggregate information for a number of national economies that are reasonably well represented by the D\&B data (see Appendix \ref{sub:country_compare}). This reduced representivity compared to the US will result in some uncertainty about the accuracy of country-level relatedness matrices.

Another drawback of using D\&B data is that, although they provide a fairly accurate account of the level of economic activity in a particular year, observed changes in economic activity tend to be quite noisy due to the fact that the information in the database is not updated in a uniform way \citep{neumark2007employment,crane2019business}. To limit such concerns, we will calculate growth rates over the longest possible time period. In spite of these shortcomings, the micro-level character of the data, the record of ownership ties and the fact that industry classifications do not change over time or across geography make the dataset uniquely suited for our purposes.

\section{Specification search}

\subsection{Setting up the search grid}\label{sec:gridsetup}
\begin{sidewaysfigure} 
	\begin{center}
	\includegraphics[scale=.82]{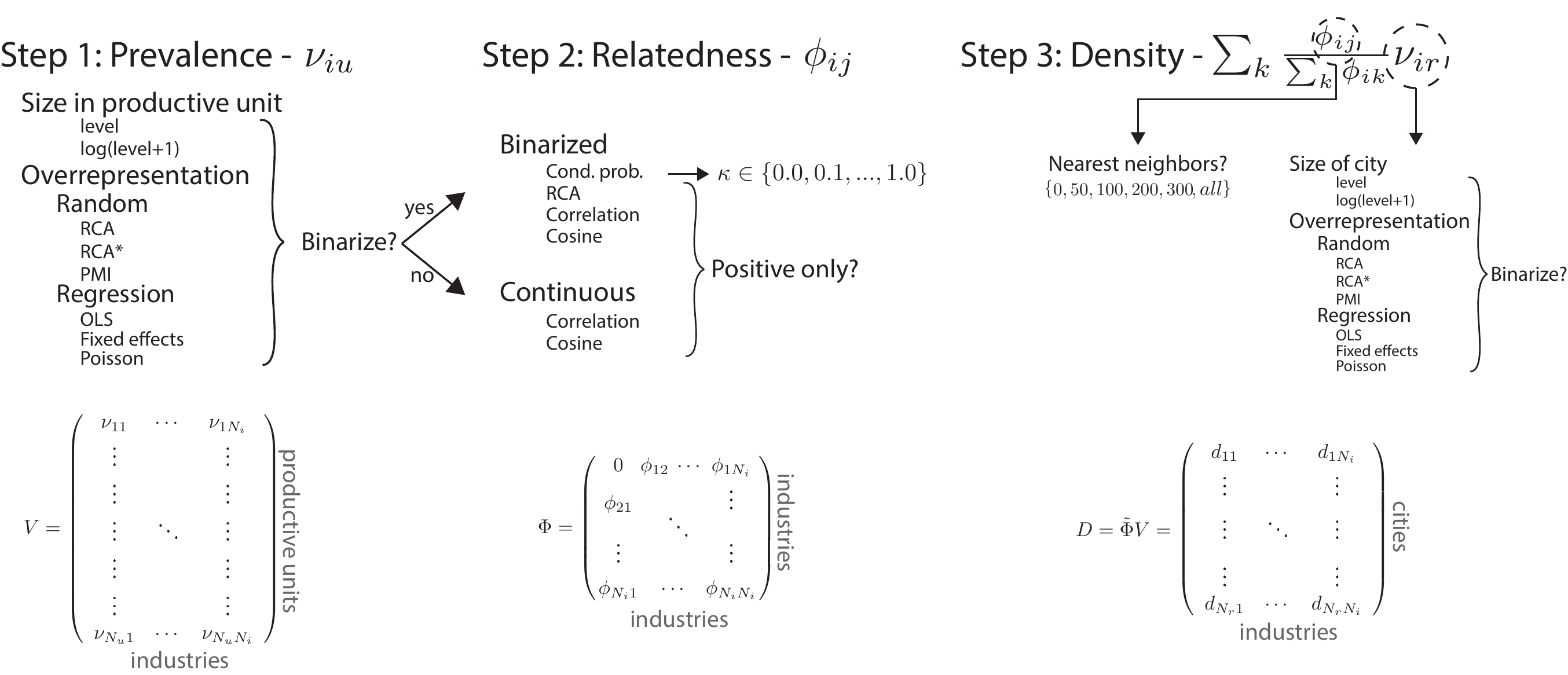}
    \caption{Constructing the specification grid.}
    \label{fig:grid}
	\end{center}
\end{sidewaysfigure}

Researchers need to make a number of choices when analyzing the principle of relatedness. We summarize these choices in Fig. \ref{fig:grid}. First, we need to choose the type of \emph{productive unit} in which industries can coincide. That is, we need to determine at which level of aggregation we capture  economies of scope. 

The D\&B data allow us to identify the industrial portfolios of four types of productive units: establishments, firms, regions and countries.\footnote{When we construct the industrial portfolios of firms, we do not use all industries listed by their establishments, but only establishments' primary industry codes (see Appendix \ref{sub:prevalence}). This way, we avoid that industries that coincide in establishments by construction also do so in firms.}  As a regional unit, we use US Core-Based Statistical Areas (CBSAs, i.e.,  metropolitan and micropolitan areas, henceforth ``cities''). 

As we move from establishments to firms, cities and countries, industrial portfolios start reflecting a widening range of economies of scope. As a consequence, the notion of relatedness changes. For instance, establishments are likely to combine activities that share similar inputs, technologies or skills \citep{neffke2011regions}. Firms can harness additional economies of scope across their establishments, by pooling managerial, marketing, sales or other organizational processes \citep{teece1994understanding}. At the level of cities, industries may coagglomerate to share pools of specialized labor and suppliers, or physical and institutional infrastructure \cite[see, e.g.,][]{ellison2010causes,diodato2018industries}. In countries, the potential sources of economies of scope widen further to include climatic conditions and macro-level institutions such as intellectual property rights regimes or sophisticated financial markets \citep{rajan1998financial}. 

However, moving to higher level productive units is not costless: it increases the number of spurious and indirect relations between industries. For instance, ski-resorts exhibit few economies of scope with hydroelectric power plants. That is why we do not observe firms that specialize in both activities, let alone that combine them in one and the same economic establishment. Yet, these industries do often colocate  in the same regions. Such combinations, which are more likely to be found in the industrial portfolios of higher-order productive units, confound relatedness as an estimate of economies-of-scope. We will study these issues in more depth in section \ref{sec:results_relatedness}. 

Once we have picked a  type of productive unit, we need to construct three interrelated quantities. The \emph{prevalence} of an economic activity (step 1 of Fig. \ref{fig:grid}), the \emph{relatedness} of pairs of industries (step 2) and the \emph{density} of related activity around an industry in a local economy (step 3). To facilitate calculations, we collect our data into matrices at the bottom of Fig. \ref{fig:grid}.

Prevalence, $\nu_{iu}$, expresses how strongly represented an industry $i$ is in productive unit $u$. The simplest way to capture this is by the amount of employment unit $u$ has in industry $i$. A more commonly used metric is the unit's revealed comparative advantage (RCA) in the industry \citep[e.g.][]{hidalgo2007product}. In appendix \ref{app:grid}, we list a number of plausible alternatives that mainly differ in the benchmark against which they assess how large an industry's presence in a productive unit is. 

When it comes to measuring relatedness between industries, our grid relies on a class of so-called outcome-based relatedness metrics. These metrics focus on the imprint that relatedness leaves on the behavior of economic actors.\footnote{\cite{neffke2013skill} distinguish between resource-based and outcome-based relatedness measures. Resource-based measures define relatedness as the extent to which industries utilize the same resources or inputs. Examples are relatedness measures based on input-output tables \citep[e.g.,][]{fan2000measurement}, occupational employment vectors \citep[e.g.,][]{farjoun1994beyond} or labor flows \citep{neffke2013skill}.} 

Outcome-based measures have been justified by reference to the survivor principle \citep{teece1994understanding}. Accordingly, the fact that we observe that two industries often coincide in the same productive units means that this combination is economically viable. Outcome-based measures have two advantages. First, there are many ways in which activities may be related. For instance, industries may share the same human capital requirements, use similar resources or be part of the same value chains. Outcome-based measures summarize all these linkages in a single metric that implicitly puts most weight on the linkages that matter most in the portfolio decisions of economic actors. Second, outcome-based measures can be derived from the same data that are used to study industrial growth or diversification. For instance, in economic geography, they can be derived from data that describe the industry mix of cities. As a consequence, outcome-based relatedness measures are by far the most common in the literature. 

Most outcome-based relatedness matrices are based on counts of how often two industries co-occur in the same productive units. These counts are typically normalized to account for how often we would expect the industries to co-occur merely by chance. We explore several metrics that differ mainly in these normalizations:  correlations, cosine distances, conditional probabilities, and RCA-like transformations. We explain the differences and similarities between these methods from a mathematical and conceptual point of view in Appendix \ref{sec:relatedness}. In addition,  we distinguish between approaches that use continuous prevalence information and those that binarize prevalence information. The latter consider that an industry is \emph{present} in a productive unit whenever its prevalence exceeds some threshold value.

Furthermore, some authors \citep[e.g., ][]{muneepeerakul2013urban} have argued that positive relatedness (industries that appear together more often than by chance) is qualitatively different from negative relatedness (industries that coincide less than by chance). We explore this by using not only the full relatedness matrices, but also versions in which  all negative elements are set to zero. 

Finally, in step 3, we need to assess not just the relatedness between two industries, but also how related an industry is to a city's entire portfolio of industries. To do so, we use a measure that \cite{hidalgo2007product} dubbed an industry's ``density'' in the city. The density of industry $i$ in city $r$, $d_{ir}$, is defined as the average prevalence of all other industries, $j$, in $r$, weighted by their relatedness to $i$:

\begin{equation} \label{eq:density}
    d_{ir} = \sum \limits_{j \neq i \in J_{i}} \frac{\phi_{ij}}{\sum \limits_{k \neq i \in J_{i}}\phi_{ik}} \nu_{jr}.
\end{equation}
where $J_{i}$ is the set of $k$ most closely related ``neighboring'' industries to industry $i$,  $\phi_{ij}$ the relatedness between industry $i$ and $j$ and $\nu_{jr}$ industry $j$'s prevalence in city $r$.\footnote{If density is used to predict employment \emph{levels}, the contribution of the industry itself to the sum in eq. (\ref{eq:density}) would render the regression analysis of section \ref{sec:densregression} circular. When instead predicting growth, we will capture the industry's own contribution by adding a mean-reversion term. Therefore, we exclude industry $i$ from its own neighborhood in eq. (\ref{eq:density}).} Once again, we can either use continuous prevalence information as weights or binary information on whether or not an industry is present in a city.

This procedure leads to various ways to quantify prevalence, relatedness and density. Combining all alternatives of each step in Fig. \ref{fig:grid}, we arrive at 32,480 different specifications.

\subsection{Estimation}\label{sec:densregression}
To evaluate each specification, we study its performance in two prediction tasks: out-of-sample prediction of employment levels and out-of-sample prediction of employment growth. For the first task, we estimate the following regression model: 
\begin{equation}\label{eq:dens_reg_levels}
\log E_{irt} =  \delta d_{irt} + \beta_1 \log E_{it} + \beta_2 \log E_{rt} + \epsilon_{irt}, 
\end{equation}
where $d_{irt}$ represents the density of industry $i$ in city $r$ at time $t$,\footnote{Because $d_{irt}$ can have a highly skewed distribution, we log-transform density in specifications where an industry's prevalence in a city is based on RCAs or on raw employment counts.} $E_{irt}$ the employment of industry $i$ in city $r$ at time $t$, $E_{it}$ the employment in industry $i$ at time $t$ and $E_{rt}$ the employment in city $r$ at time $t$. $\epsilon_{irt}$ is a residual.

For the second task, we estimate the following growth model:
\begin{equation}\label{eq:dens_reg_intensive}
\Delta \log E_{irt}  = \gamma \log E_{irt} +  \delta d_{irt} + \beta_1 \log E_{it} + \beta_2 \log E_{rt} + \epsilon_{irt} \text{  if $E_{irt}>0, E_{irt+\theta}>0$;}
\end{equation}
where $\theta$ is the time-horizon over which growth is measured and $\gamma$ captures mean reversion effects.\footnote{These effects -- which quantify the influence of the size of an industry in the base year -- have sometimes been interpreted as local competition effects \citep[e.g. ][]{delgado2014clusters}. However, negative mean reversion effects also arise as statistical artefacts if $E_{irt}$ is measured with noise or has an idiosyncratic element.} Accounting for the latter is crucial, because $E_{irt}$ and $d_{irt}$ are typically strongly and positively correlated. Therefore, failing to account for (negative) mean reversion effects will lead to an underestimation of the (typically positive) effect of density.\footnote{Different studies add different control variables. Apart from controls for the overall size of the region and of the industry, some studies add aggregate growth rates of regions and industries -- what \cite{hausmann2021implied} call \emph{radial growth}. Another common specification adds industry and region dummies. Note, however, that both sets of control variables assume information that is not available in a forecasting exercise: radial growth explicitly assumes that aggregate future growth rates are available and region and industry fixed effects make the same assumption implicitly.} Note that we only evaluate performance for growth at the intensive margin. That is, we only look at growth rates for local industries that exist in both 2011 and 2019. This simplifies the analysis because it avoids values of $\log(0)$ in the dependent variable and mean reversion term.

To evaluate the models of eqs (\ref{eq:dens_reg_levels}) and (\ref{eq:dens_reg_intensive}), we first divide the dataset into a train and a test sample. The train sample is used to construct density and control variables, as well as to fit the model's parameters. To make it computational feasible to run millions of regression analyses, we fit these models using Ordinary Least Squares (OLS). Next, predictive performance is evaluated on the test sample. We repeat this procedure 100 times to arrive at an average model fit for each candidate specification, expressed as the model's out-of-sample $R^2$, as well as confidence bands around this average. Finally, as a benchmark, we  estimate models without density terms and with density terms that are based on (symmetric) relatedness matrices whose elements are drawn at random from a uniform distribution. 

\section{Results} \label{sec:results}
\subsection{Grid search} \label{sec:results_grid}
\begin{figure} 
	\begin{center}
	\includegraphics[scale=.85]{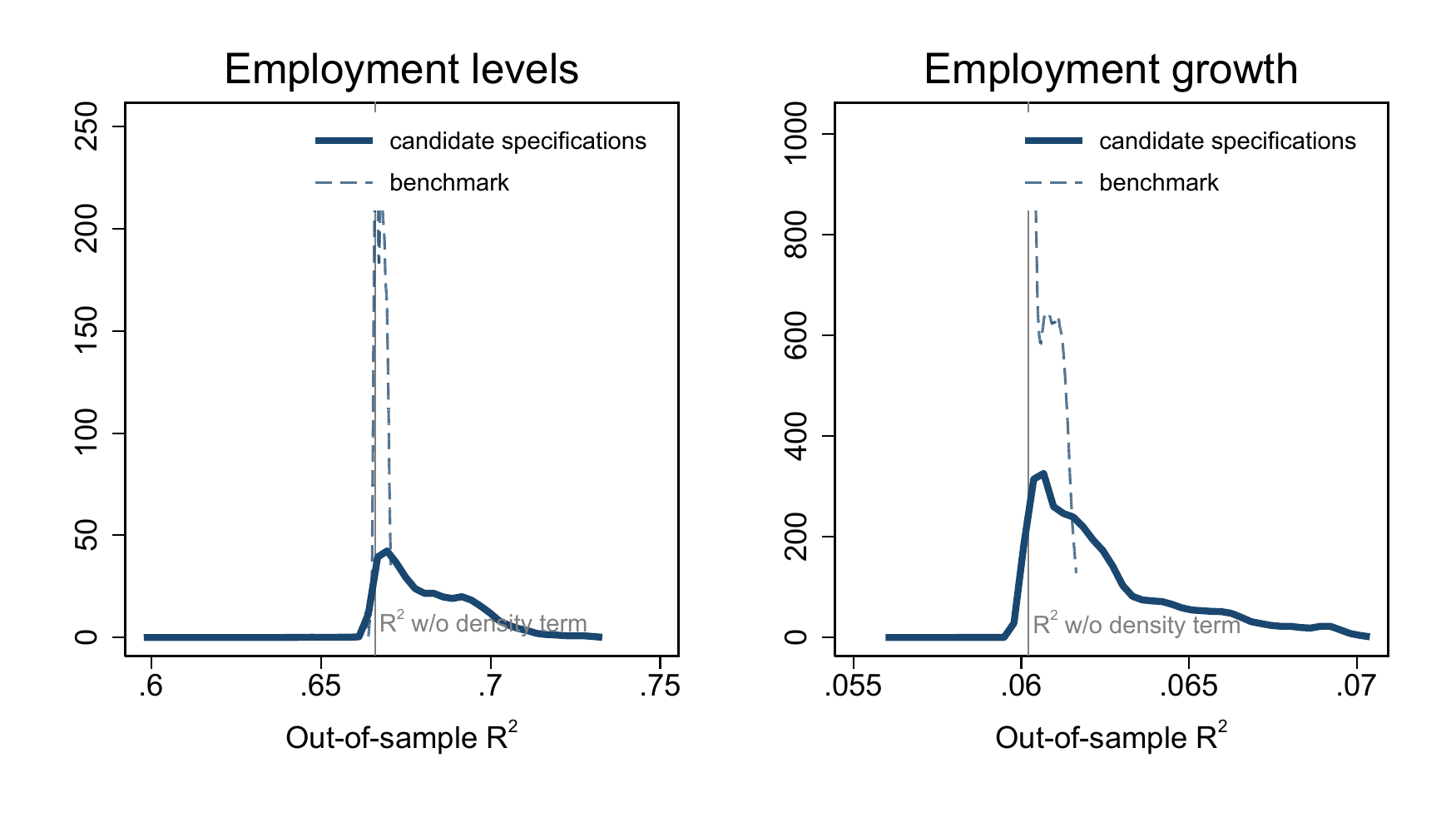}
    \mycaption{Kernel density plots of out-of-sample $R^2$.}{Left: prediction of employment levels; right: prediction of employment growth. Vertical gray lines: $R^2$ of specifications of eqs (\ref{eq:dens_reg_levels}) and (\ref{eq:dens_reg_intensive}) without density ($d_{irt}$) term. \emph{Candidate specifications} shows the distribution of the $R^2$ in the specification grid. \emph{Benchmark} shows the distribution for runs in which density terms are based on random proximity matrices.}\label{fig:R2_density}
	\end{center}
\end{figure}

Fig. \ref{fig:R2_density} shows distributions of model performance across specifications. The benchmark specification without density term is marked by vertical lines and dashed curves show benchmark performance distributions for specifications with random relatedness matrices. 

The majority of specifications corroborate the principle of relatedness: 88\% of candidate specifications outperform the median random benchmark in predicting employment levels and 78\% in predicting employment growth. However, there is substantial heterogeneity across specifications. For specifications in the \nth{99} performance percentile, adding a density term raises the $R^2$  by about 10\% when predicting employment levels and by about 16\% when predicting employment growth. In the median specification, the density term raises the $R^2$ of these models only by between 2 and 3\%. 

Appendix \ref{app:Shapley_value} describes which specification elements matter most for a model's performance. The most important aspect of the specification grid turns out to be how we define density and, in particular, how we define the prevalence of an industry in a city. Also the choice of the productive unit has a large impact on the quality of growth predictions. Of least importance is how many neighbors are used when calculating density variables. 

Table \ref{tab:bestspec} shows the two top specifications in terms of predictive validity. The first column uses the prediction of employment levels as a criterion, the second the prediction of employment growth. Apart from relying on the same productive unit for relatedness calculations, the selected specifications have little in common. Moreover, their performance is not very robust: the specification that predicts employment levels best, fails to do well in predicting employment growth, and vice versa. This suggests that the performance criteria are too noisy to confidently select an optimal specification from our grid.

\begin{table}
    \begin{center}
        \begin{tabular}{lccc|c} 
  & \multicolumn{2}{c}{\textbf{prediction task}} & & \textbf{baseline} \\
 & $ \log(\text{emp.}) $ & $ \Delta\log(\text{emp.}) $ & & \\
\hline
& \multicolumn{4}{c}{\emph{Relatedness}} \\
\hline
entity & City & City & & n.a. \\
prevalence & resid. (OLS) & RCA & &  n.a. \\
binarized & no & yes & & n.a. \\
metric & cosine dist. & RCA* & & n.a. \\
$ \kappa $ & n.a. & n.a. & & n.a. \\
truncation & yes & yes & & n.a. \\
\hline
& \multicolumn{4}{c}{\emph{Density}} \\
\hline
prevalence & resid. (OLS) & RCA & & n.a. \\
binarized & yes & yes & & n.a. \\
\# neighbors & 300 & 100 & &  n.a. \\
\hline
& \multicolumn{4}{c}{\emph{Out-of-sample} $ R^2$ } \\
\hline
$ \log(\text{emp.}) $ &  0.770 & 0.710 & & 0.666  \\
$ \Delta\log(\text{emp.}) $ & 0.063 & 0.076 & & 0.060 \\
\hline\hline
\end{tabular}

    \end{center}
    \mycaption{Top specifications.}{Specifications with the highest out-of-sample $R^2$ for employment levels (first column), for employment growth regressions (second column) and without density term (third column).}\label{tab:bestspec}
\end{table}

\subsection{Robust performance}\label{sec:robust}
To draw reliable conclusions, we explore which specification choices are robustly associated with good performance. We will call a specification \emph{robust} if it ranks among the best 10\% of specifications in both prediction tasks. Next, we count how often each choice is used in this set of robust specifications. 

Tables \ref{tab:relatedness} and \ref{tab:density} report the results of this exercise. They display for each specification characteristic the share of specifications that yield robust performance. These shares can be interpreted as the likelihood that a randomly chosen specification that uses a given characteristic preforms robustly. Columns correspond to the types of  productive units that were used to measure inter-industry relatedness. Furthermore, the tables are split into two parts, one using binary (presence) information and one using continuous (prevalence) information to construct relatedness matrices (Table \ref{tab:relatedness}) or density variables (Table \ref{tab:density}).  

\begin{table}
    \begin{center}
        \begin{small}        \begin{tabular}{lrrrrr} 
\multicolumn{6}{c}{\textbf{(BINARY) PRESENCES}} \\
\hline
\emph{PREVALENCE} & Cntry & City & Firm & Estab. & Total \\
\hline
RCA/RCA*/PMI & 0.0\% & 3.0\% & 0.6\% & 0.1\% & 0.9\% \\ 
resid. (POI) & 0.0\% & 3.0\% & 0.6\% & 0.0\% & 0.9\% \\ 
emp./log(emp.) & 0.0\% & 0.0\% & 0.6\% & 0.1\% & 0.2\% \\ 
\hline 
Total & 0.0\% & 5.9\% & 1.9\% & 0.2\% & \textbf{2.0\%} \\ 
\hline
\emph{METRIC} & Cntry & City & Firm & Estab. & Total \\
\hline
RCA* & 0.0\% & 1.7\% & 0.0\% & 0.0\% & 0.4\% \\ 
Pearson corr. & 0.0\% & 1.3\% & 0.1\% & 0.0\% & 0.3\% \\ 
cosine dist. & 0.0\% & 0.0\% & 0.3\% & 0.2\% & 0.1\% \\ 
$ \kappa $ = 0.0 & 0.0\% & 1.2\% & 0.3\% & 0.0\% & 0.4\% \\ 
... 0.1 & 0.0\% & 0.9\% & 0.3\% & 0.0\% & 0.3\% \\ 
... 0.2 & 0.0\% & 0.5\% & 0.3\% & 0.0\% & 0.2\% \\ 
... 0.3 & 0.0\% & 0.2\% & 0.2\% & 0.0\% & 0.1\% \\ 
... 0.4 & 0.0\% & 0.0\% & 0.2\% & 0.0\% & 0.1\% \\ 
... 0.5 & 0.0\% & 0.0\% & 0.2\% & 0.0\% & 0.0\% \\ 
\hline 
Total & 0.0\% & 5.9\% & 1.9\% & 0.2\% & \textbf{2.0\%} \\ 
\hline\hline
\end{tabular}

        \begin{tabular}{lrrrrr} 
\\
\multicolumn{6}{c}{\textbf{(CONTINUOUS) PREVALENCES}} \\
\hline
\emph{PREVALENCE} & Cntry & City & Firm & Estab. & Total \\
\hline
RCA* & 0.0\% & 2.0\% & 0.2\% & 0.2\% & 0.6\% \\ 
resid. (OLS) & 0.0\% & 0.2\% & 1.3\% & 0.1\% & 0.4\% \\ 
PMI & 0.0\% & 0.4\% & 0.0\% & 0.0\% & 0.1\% \\ 
emp. & 0.0\% & 0.0\% & 0.0\% & 0.1\% & 0.0\% \\ 
resid. (POI) & 0.0\% & 0.1\% & 0.0\% & 0.0\% & 0.0\% \\ 
RCA & 0.0\% & 0.0\% & 0.0\% & 0.1\% & 0.0\% \\ 
log(emp.) & 0.0\% & 0.0\% & 0.0\% & 0.1\% & 0.0\% \\ 
\hline 
Total & 0.0\% & 2.8\% & 1.4\% & 0.6\% & \textbf{1.2\%} \\ 
\hline
\emph{METRIC} & Cntry & City & Firm & Estab. & Total \\
\hline
Pearson corr. & 0.0\% & 2.7\% & 0.6\% & 0.0\% & 0.8\% \\ 
cosine dist. & 0.0\% & 0.1\% & 0.8\% & 0.6\% & 0.4\% \\ 
\hline 
Total & 0.0\% & 2.8\% & 1.4\% & 0.6\% & \textbf{1.2\%} \\ 
\hline\hline
\end{tabular}

       \end{small} 
    \end{center}
    \mycaption{Definition of relatedness.}{}\label{tab:relatedness}
\end{table}

\begin{table}
    \begin{center}
       \begin{small}
        \begin{tabular}{lrrrrr} 
\multicolumn{6}{c}{\textbf{(BINARY) PRESENCES}} \\
\hline
\emph{PREVALENCE} & Cntry & City & Firm & Estab. & Total \\
\hline
RCA/RCA*/PMI & 0.0\% & 2.1\% & 0.1\% & 0.4\% & 0.7\% \\ 
resid. (POI) & 0.0\% & 2.1\% & 0.1\% & 0.4\% & 0.7\% \\ 
resid. (OLS) & 0.0\% & 0.0\% & 0.4\% & 0.0\% & 0.1\% \\ 
\hline 
Total & 0.0\% & 4.3\% & 0.7\% & 0.8\% & \textbf{1.4\%} \\ 
\hline
\emph{TRUNCATION} & Cntry & City & Firm & Estab. & Total \\
\hline
positive & 0.0\% & 1.7\% & 0.3\% & 0.0\% & 0.5\% \\ 
all & 0.0\% & 0.9\% & 0.3\% & 0.8\% & 0.5\% \\ 
n.a. & 0.0\% & 1.6\% & 0.1\% & 0.0\% & 0.4\% \\ 
\hline 
Total & 0.0\% & 4.3\% & 0.7\% & 0.8\% & \textbf{1.4\%} \\ 
\hline
\emph{\# NEIGHBORS} & Cntry & City & Firm & Estab. & Total \\
\hline
50 & 0.0\% & 1.3\% & 0.3\% & 0.5\% & 0.5\% \\ 
100 & 0.0\% & 1.2\% & 0.1\% & 0.3\% & 0.4\% \\ 
200 & 0.0\% & 0.8\% & 0.1\% & 0.0\% & 0.2\% \\ 
300 & 0.0\% & 0.5\% & 0.1\% & 0.0\% & 0.1\% \\ 
415 & 0.0\% & 0.5\% & 0.1\% & 0.0\% & 0.1\% \\ 
\hline 
Total & 0.0\% & 4.3\% & 0.7\% & 0.8\% & \textbf{1.4\%} \\ 
\hline\hline
\end{tabular}

        \begin{tabular}{lrrrrr} 
\\
\multicolumn{6}{c}{\textbf{(CONTINUOUS) PREVALENCES}} \\
\hline
\emph{PREVALENCE} & Cntry & City & Firm & Estab. & Total \\
\hline
resid. (OLS) & 0.0\% & 1.3\% & 2.3\% & 0.0\% & 0.9\% \\ 
RCA* & 0.0\% & 1.1\% & 0.1\% & 0.0\% & 0.3\% \\ 
resid. (POI) & 0.0\% & 0.9\% & 0.0\% & 0.0\% & 0.2\% \\ 
PMI & 0.0\% & 0.9\% & 0.0\% & 0.0\% & 0.2\% \\ 
log(emp.) & 0.0\% & 0.7\% & 0.1\% & 0.0\% & 0.2\% \\ 
emp. & 0.0\% & 0.6\% & 0.0\% & 0.0\% & 0.2\% \\ 
\hline 
Total & 0.0\% & 5.5\% & 2.5\% & 0.0\% & \textbf{2.0\%} \\ 
\hline
\emph{TRUNCATION} & Cntry & City & Firm & Estab. & Total \\
\hline
positive & 0.0\% & 2.4\% & 0.4\% & 0.0\% & 0.7\% \\ 
all & 0.0\% & 0.9\% & 0.2\% & 0.0\% & 0.3\% \\ 
n.a. & 0.0\% & 2.3\% & 1.8\% & 0.0\% & 1.0\% \\ 
\hline 
Total & 0.0\% & 5.5\% & 2.5\% & 0.0\% & \textbf{2.0\%} \\ 
\hline
\emph{\# NEIGHBORS} & Cntry & City & Firm & Estab. & Total \\
\hline
100 & 0.0\% & 1.4\% & 0.5\% & 0.0\% & 0.5\% \\ 
50 & 0.0\% & 1.3\% & 0.4\% & 0.0\% & 0.4\% \\ 
200 & 0.0\% & 1.1\% & 0.6\% & 0.0\% & 0.4\% \\ 
415 & 0.0\% & 0.9\% & 0.5\% & 0.0\% & 0.3\% \\ 
300 & 0.0\% & 0.8\% & 0.5\% & 0.0\% & 0.3\% \\ 
\hline 
Total & 0.0\% & 5.5\% & 2.5\% & 0.0\% & \textbf{2.0\%} \\ 
\hline\hline
\end{tabular}

        \end{small} 
    \end{center}
    \mycaption{Definition of density.}{}\label{tab:density}
\end{table}

The overall shares of robust specifications are provided in bold. The likelihood that a random specification is robust is quite low: 2.0\% for specifications with binary (co-occurrence-based) relatedness matrices and 1.2\% for specifications with continuous (co-prevalence-based) relatedness matrices. 

\paragraph*{Preferred specifications.} In appendix \ref{app:recommendations}, we provide recommendations for how to avoid  poor predictive performance, listing a number of general lessons about how \emph{not} to construct relatedness and density metrics. Here, instead, we analyze two specifications that, in light of Tables \ref{tab:relatedness} and \ref{tab:density}, are \emph{a priori} expected to work particularly well. The first specification uses a relatedness matrix based on city-level binarized co-occurrence information. The second uses a relatedness matrix based on firm-level continuous co-prevalence information. As points of reference, we also consider the specification in \cite{hidalgo2007product}, once based on country-level co-occurrences as used in  \citeauthor{hidalgo2007product}'s original study and once based on city-level co-occurrences. The latter are frequently used in the economic geography literature \citep{boschma2013emergence,montresor2017regional,zhu2017jump}. 

\begin{table}[t!]
    \begin{center}
        \begin{tabular}{lccccc|ccccc} 
 & \multicolumn{5}{c|}{\textbf{Preferred specifications}} & \multicolumn{5}{c}{\textbf{Hidalgo et al. (2007)}} \\
 & \multicolumn{2}{c}{\textbf{binary}} & & \multicolumn{2}{c|}{\textbf{continuous}} &  \multicolumn{2}{c}{\textbf{country}} & &  \multicolumn{2}{c}{\textbf{city}} \\
\hline
 & \multicolumn{10}{c}{\emph{Relatedness}}  \\
\hline
entity & \multicolumn{2}{c}{city} & & \multicolumn{2}{c|}{Firm} & \multicolumn{2}{c}{Country} & & \multicolumn{2}{c}{city} \\
prevalence & \multicolumn{2}{c}{RCA} & & \multicolumn{2}{c|}{resid. (OLS)} & \multicolumn{2}{c}{RCA} & & \multicolumn{2}{c}{RCA} \\
binarized & \multicolumn{2}{c}{yes} & & \multicolumn{2}{c|}{no} & \multicolumn{2}{c}{yes} & & \multicolumn{2}{c}{yes} \\
metric & \multicolumn{2}{c}{RCA*} &  & \multicolumn{2}{c|}{cosine dist.} & \multicolumn{2}{c}{cond. prob.} & & \multicolumn{2}{c}{cond. prob.} \\
$ \kappa $ & \multicolumn{2}{c}{n.a.} & & \multicolumn{2}{c|}{n.a.} & \multicolumn{2}{c}{0.0} & & \multicolumn{2}{c}{0.0} \\
truncation & \multicolumn{2}{c}{yes} & & \multicolumn{2}{c|}{yes} & \multicolumn{2}{c}{n.a.} & & \multicolumn{2}{c}{n.a.} \\
\hline
& \multicolumn{10}{c}{\emph{Density}} \\
\hline
prevalence & \multicolumn{2}{c}{resid. (OLS)} & & \multicolumn{2}{c|}{resid. (OLS)} & \multicolumn{2}{c}{RCA} & & \multicolumn{2}{c}{RCA} \\
binarized & \multicolumn{2}{c}{no} & & \multicolumn{2}{c|}{no} & \multicolumn{2}{c}{yes} & & \multicolumn{2}{c}{yes} \\
\# neighbors & \multicolumn{2}{c}{100} & & \multicolumn{2}{c|}{100} & \multicolumn{2}{c}{415} & & \multicolumn{2}{c}{415} \\
\hline
& \multicolumn{10}{c}{\emph{Out-of-sample performance}}  \\
\hline
$ R^2 $ - levels & 0.707 & (0.950) & & 0.730 & (0.989) & 0.685 & (0.623) & & 0.704 & (0.929) \\
$ R^2 $ - growth & 0.069 & (0.974) & & 0.068 & (0.955) & 0.062 & (0.508) & & 0.067 & (0.926) \\
\hline\hline
\end{tabular}

    \end{center}
    \mycaption{Preferred specification.}{Left panel: specifications with binary and continuous prevalence used to construct relatedness matrices. Right panel: specifications in \cite{hidalgo2007product} using industry combinations observed within countries and cities. The bottom rows show the specification's performance in terms of out-of-sample $R^{2}$, with the corresponding percentile rank position in parentheses.}\label{tab:specapriori}
\end{table}

Table \ref{tab:specapriori} shows that both preferred specifications work remarkably well, ranking among the top 5\% of specifications in terms of predicting employment and employment growth.  \citeauthor{hidalgo2007product}'s specification performs only marginally worse, as long as it uses city colocation information to calculate relatedness.

\section{Inside the principle of relatedness}\label{sec:substantivefindings}
A core aspect of each empirical specification is its relatedness matrix. These matrices depend somewhat on technical details, such as normalizations and proximity metrics (see Fig. \ref{fig:corrplot} of Appendix \ref{app:correlation_relmat}), but the most pronounced differences are due to differences in the productive units in which industry co-occurrences are recorded. Studying these differences offers an opportunity to learn more about the inner workings of the principle of relatedness. To do so, we focus on three aspects: (1) the degree to which matrices yield well-delineated clusters of industries, (2) the drivers of relatedness, and (3) their performance in predicting future growth.  

\subsection{Industry spaces} \label{sec:results_relatedness}

Relatedness matrices are often visualized as networks, or \emph{industry spaces} \citep[e.g., ][]{hidalgo2007product}. These networks can be used to identify clusters of related industries and how to do so is an active field of research \citep{delgado2016defining,o2022modular}. Here, we explore how well different matrices lend themselves for this task, focusing on the relatedness matrices in our preferred specifications. 

\subsubsection*{Delineating clusters of industries}

Appendix \ref{APP:industry_spaces} displays two industry spaces for each productive unit, one for our preferred binary and one for our preferred continuous specification. Some industry spaces exhibit more structure than others. Although deciding how ``structured'' a network is is somewhat subjective, we can quantify an aspect of this by determining how easy it is to identify communities in the network.

To do so, we calculate  the \emph{modularity}  and the \emph{effective number of communities} for each industry space. Modularity quantifies to what extent a network consists of easily separable communities. It is defined as the fraction of ties that form between industries belonging to the same community, minus the fraction of such ties that we would have expected, had links formed at random. A high modularity score thus means that most links form between industries in the same community, as opposed to across communities, i.e., that the industry space is composed of distinct, well-delineated clusters.\footnote{Because relatedness matrices are too dense for network analysis, we use the truncated networks of Figs \ref{fig:indspace_binary} and \ref{fig:indspace_continuous}, not of the full relatedness matrices. To calculate modularity, we rely on the Louvain algorithm \citep{blondel2008fast} to compute the best community division and then calculate the corresponding modularity scores. The random benchmark used in these calculations preserves node degrees, but randomly rewires connections.} The effective number of communities is a so-called \emph{Hill number} \citep{hill1973diversity}. It is calculated as $e^H$, where $H$ represents the entropy of the communities' size distribution. It reflects how many communities of equal size would be needed to arrive at the same observed entropy.

\begin{figure} 
	\begin{center}
    	\includegraphics[width=.75\textwidth]{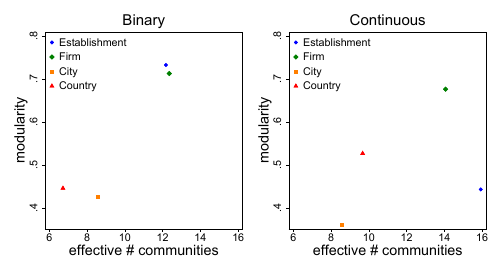}
        \mycaption{Modularity and effective number of communities.}{The vertical axis depicts the modularity and the horizontal axis the effective number of communities -- defined as $e^H$, where $H$ denotes the entropy of the size distribution of the communities -- for different relatedness matrices. Left panel uses our preferred specification for binary presence data to construct the relatedness matrices, right panel for continuous prevalence data.}\label{fig:modularity}
	\end{center}
\end{figure}

Results are shown in Fig. \ref{fig:modularity}. Industry spaces based on firm-level portfolios exhibit the clearest community structures, combining a modest number of distinct communities with a high degree of modularity. In contrast, cities give rise to the most cluttered industry spaces with only few not very well-delineated communities. The industry spaces for establishments and countries are somewhere in between these two extremes.

\subsubsection*{Conformance to sectoral classification}
Do industries that belong to the same higher-level sectors tend to be more related? To study this, we define the \emph{SIC relatedness} between two industries as the number of leading digits their SIC codes shares. Given that we work with 3-digit industries, SIC relatedness can be 0, 1 or 2. If relatedness is particularly high between industries in the same sector but not across sectors, we will say that the relatedness matrix conforms closely to the classification system's sectoral boundaries. Because of our familiarity with classical sector designations, such matrices tend to be more intuitive.

\begin{figure} 
	\begin{center}
    	\includegraphics[width=.75\textwidth]{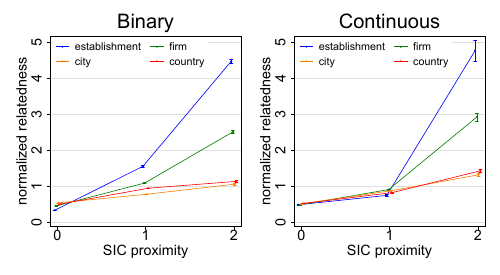}
        \mycaption{Agreement with the industrial classification system.}{The vertical axis depicts the average relatedness and its 95\% confidence interval between two industries based on our preferred binary (left) or continuous (right) specification. To ensure comparability, links are normalized by the sum total of elements in each relatedness matrix, multiplied by 100,000 for graphical convenience. The horizontal axis depicts industries' SIC relatedness, where 0 means that the industries belong to different sectors.}\label{fig:SIC_relatedness}
	\end{center}
\end{figure}

Fig. \ref{fig:SIC_relatedness} plots the average relatedness for industry pairs at different levels of SIC relatedness. The binary and continuous specifications conform about equally closely to the higher-level sector-boundaries. However, there are pronounced differences between productive units. The closest agreement with the classification system is achieved by relatedness matrices based on the industrial portfolios of establishments. The next closest agreement relies on firm portfolios. In contrast, industrial portfolios of cities and countries yield relatedness matrices that align much less with the sectoral classification. As a consequence, the establishment- and firm-level relatedness matrices yield quite intuitive clusters. In contrast, relatedness based on city- or country-level colocation patterns often cuts across sector boundaries, suggesting connections that are more surprising at best and puzzling at worst.

\subsubsection*{Drivers of  relatedness}
In section \ref{sec:gridsetup}, we speculated that different types of productive units harness different types of economies of scope. To explore this, we regress the number of co-occurrences between industries on measures of the strength of value chain linkages and human capital similarities. To quantify value-chain linkages, we cast US input-output tables \citep{ingwersen2022useeio} into the SIC 3-digit classification. Next, we symmetrize these tables by using the maximum of the input coefficients between industries $i$ and $j$, $a_{ij}$, in either direction: 

\begin{equation*}\label{eq:valuechain}
\phi^{vc}_{ij} = \max(a_{ij}, a_{ji})    
\end{equation*}

To quantify human capital similarities, we rely on the US Bureau of Labor Statistics' Occupational Employment and Wage Statistics for 2011.\footnote{Retrieved from \url{www.bls.gov/oes}.} We again  transform industry codes into 3-digit SIC industries. The human capital similarity between industry $i$ and $j$ is now calculated as:

\begin{align*}\label{eq:labor}
\pi_{oi} =  I \left( \frac{E_{oi}/E_{i}}{E_{o}/E} > 1 \right),\\
c_{ij} = \pi_{oi} \pi_{oj}, \\
RCA_{ij} = \frac{c_{ij}}{c_{i}c_{j}}c, \\
\phi^{hc}_{ij} = \frac{RCA_{ij}}{RCA_{ij}+1}
\end{align*}
where $I(.)$ is an indicator function that evaluates to 1 if its argument is true and to 0 otherwise, $E_{oi}$ is the employment of occupation $o$ in industry $i$, $E_o$ and $E_i$ the employment of occupation $o$ and industry $i$ in the US and $E$ the overall employment in the US. In words, we first determine which occupations are overrepresented in which industries. Next, we count how many occupations are overrepresented in industry $i$ \emph{and} $j$. Finally, we normalize these occupational co-occurrences using an RCA-style metric, that we cast between 0 and 1 to reduce skew, mimicking the steps described in eqs (\ref{eq:RCA}), (\ref{eq:RCA_T}) and (\ref{eq:cooc_bin_norm}):

\begin{table}[t!]
    \begin{center}
        \begin{tabular}{l*{4}{c}}
\hline\hline
            &\multicolumn{1}{c}{(1)}&\multicolumn{1}{c}{(2)}&\multicolumn{1}{c}{(3)}&\multicolumn{1}{c}{(4)}\\
            &\multicolumn{1}{c}{Establishment}&\multicolumn{1}{c}{Firm}&\multicolumn{1}{c}{City}&\multicolumn{1}{c}{Country}\\
\hline
$\phi^{hc}$       &       7.344\sym{***}&       2.424\sym{***}&       1.036\sym{***}&       1.114\sym{***}\\
            &      (1.07)         &      (0.24)         &      (0.00)         &      (0.01)         \\
$\phi^{vc}$          &       1.056\sym{**} &       1.044\sym{***}&       1.012\sym{***}&       1.014\sym{***}\\
            &      (0.02)         &      (0.01)         &      (0.00)         &      (0.00)         \\

\hline
\(N\)       &       85,464         &       85,029         &       85,903         &       85,903         \\
pseudo-$R^2$        &       0.784         &       0.749         &       0.860         &       0.453         \\
\hline\hline

\end{tabular}
    \end{center}
  \mycaption{Drivers of co-occurrences of industries.}{Standard errors (two-way clustered by industry) in parentheses, \sym{*}: \(p<0.05\), \sym{**}: \(p<0.01\), \sym{***}: \(p<0.001\). The table reports incidence rate ratios from a Poisson pseudo-maximum-likelihood regression of the number of co-occurrences of two industries in a productive unit on human capital similarity and value-chain linkages between the industries, controlling for two-way industry fixed effects. Coefficients are standardized by each variable's standard deviation. Co-occurrence counts refer to the number of times that two industries exhibit $RCA>1$ in the same productive unit. Columns label the type of productive unit in which co-occurrences are observed.}\label{tab:cooc_drivers}
\end{table}

Next, we regress co-occurrence counts on value chain and human capital linkages between industries. We standardize explanatory variables by mean-centering them and dividing them by their standard deviations. As a result, the reported incidence rate ratios (IRRs) express by which factor co-occurrences go up for a one-standard-deviation increase in human capital similarities or value-chain linkages. Table \ref{tab:cooc_drivers} reports results for Poisson pseudo-maximum-likelihood estimation with two-way industry fixed effects. 

The main driver of co-occurrence patterns is human capital similarities. Although this holds across all models, effects are particularly salient for co-occurrences in establishments and firms, where a one-standard-deviation increase in human capital similarity is associated with an over 7-fold increase in co-occurrences in establishments and an over 2-fold increase in firms.  In cities and countries, the link between co-occurrence patterns and labor linkages is much weaker, lifting co-occurrences by just 4\% and 11\%. This suggests that human-capital based economies of scope are core to how industries are combined in establishments and firms, but labor market externalities are much less important for how industries co-agglomerate in cities and countries. Finally, although value-chain linkages are always statistically significant, they explain comparatively little of which industries co-occur in all productive units. 


\subsubsection*{How much information does a productive unit add?}
Regardless of the productive unit in which we measure relatedness, density always has a positive and significant effect on the sizes and growth rates of local industries in our preferred specifications. But do different productive units offer redundant or complementary perspectives on relatedness?

To analyze this, we run our performance regressions with \emph{pairs} of density variables, juxtaposing the relatedness matrices associated with two different productive units. We then ask in which pairings one of the two density variables turns statistically insignificant. When this happens, the juxtaposed relatedness matrices contain redundant information.

\begin{table}[t!]
    \begin{center}
        \begin{small}
            \begin{tabular}{l*{7}{c}}
\hline\hline
\textbf{productive unit}   &\multicolumn{1}{c}{(1)}&\multicolumn{1}{c}{(2)}&\multicolumn{1}{c}{(3)}&\multicolumn{1}{c}{(4)}&\multicolumn{1}{c}{(5)}&\multicolumn{1}{c}{(6)}\\

\hline
\multicolumn{7}{c}{\textit{Effect of density on employment levels}}\\
\hline
establishment&       0.511\sym{***}&       0.318\sym{***}&                     &       0.439\sym{***}&                     &                     \\
            &    (0.0364)         &    (0.0397)         &                     &    (0.0373)         &                     &                     \\
firm&      0.0580         &                     &       0.199\sym{***}&                     &       0.306\sym{***}&                     \\
            &    (0.0367)         &                     &    (0.0426)         &                     &    (0.0328)         &                     \\
city&                     &       0.378\sym{***}&       0.462\sym{***}&                     &                     &       0.463\sym{***}\\
            &                     &    (0.0436)         &    (0.0461)         &                     &                     &    (0.0473)         \\
country&                     &                     &                     &       0.186\sym{***}&       0.289\sym{***}&       0.237\sym{***}\\
            &                     &                     &                     &    (0.0534)         &    (0.0467)         &    (0.0565)         \\
\hline
\(N\)                   & 384,705         &      384,705         &      384,705         &      384,705         &      384,705         &      384,705         \\
\(R^{2}\)   &       0.720         &       0.732         &       0.727         &       0.723         &       0.714         &       0.729         \\
\hline 
\multicolumn{7}{c}{\textit{Effect of density on employment growth}} \\
\hline 
establishment &     0.00345\sym{*}  &     0.00275         &                     &     0.00755\sym{***}&                     &                     \\
            &   (0.00174)         &   (0.00149)         &                     &   (0.00138)         &                     &                     \\
firm &     0.00576\sym{***}&                     &     0.00357\sym{*}  &                     &     0.00803\sym{***}&                     \\
            &   (0.00170)         &                     &   (0.00149)         &                     &   (0.00124)         &                     \\
city &                     &     0.00937\sym{***}&     0.00891\sym{***}&                     &                     &      0.0105\sym{***}\\
            &                     &   (0.00152)         &   (0.00155)         &                     &                     &   (0.00133)         \\
country&                     &                     &                     &    0.000895         &    0.000832         &     0.00118         \\
            &                     &                     &                     &   (0.00151)         &   (0.00139)         &   (0.00138)         \\
\hline 
\(N\)          &      245,395         &      245,395         &      245,395         &      245,395         &      245,395         &      245,395 \\
\(R^{2}\)   &       0.068         &       0.072         &       0.072         &       0.067         &       0.068         &       0.072         \\
\hline\hline
\end{tabular}    
        \end{small}
    \end{center}
  \mycaption{Redundancy and complementarity in pairs of density variables.}{Standard errors in parentheses, \sym{*}: \(p<0.05\), \sym{**}: \(p<0.01\), \sym{***}: \(p<0.001\). Upper panel: regression analysis of employment levels, controlling for industry and city size. Lower panel: regression analysis of employment growth, controlling for mean reversion effects, industry and city size. The coefficients reflect the effects of density measured using the productive units listed in the first column. All density measures are based on our preferred binarized specification. For results based on our preferred continuous specifications, see Table \ref{tab:densitypairs_con} in Appendix \ref{app:denspairs}.}\label{tab:densitypairs_bin}
\end{table}

Table \ref{tab:densitypairs_bin} shows results. To economize on space, we only display coefficients for the density variables, omitting control variables. Although firms and establishments give rise to different relatedness matrices, the information in establishments that is \emph{relevant} for our predictions is largely the same as in firms. In contrast, city-level co-occurrences offer information that is not captured in the establishment or firm level relatedness matrices. This suggests that the industrial portfolios of firms (or establishments) and of cities provide complementing views on the predicted growth trajectories of cities.\footnote{The country level also offers new information, however, only when it comes to predicting employment levels, not employment growth. Because  country-level relatedness provides only weak information about growth opportunities, we discount this evidence. These findings are corroborated when density variables are constructed with our preferred continuous instead of binary specifications (see Appendix \ref{app:denspairs}).}

\subsubsection*{Prediction versus interpretation}
Our results so far suggest an interesting tension between interpretability and predictive validity. City-level relatedness matrices tend to be top performers when it comes to out-of-sample prediction and they contain information that is not present in other relatedness matrices. Yet, city-level industry spaces are less intuitive and interpretable than firm- or establishment-level ones: they neither closely follow the delineations between traditional sectors, nor strongly reflect human capital or value-chain links.

\subsection{Sector-specific estimations}\label{sec:sectorspecific}

So far, we have studied the principle of relatedness in the economy as a whole. However, industries differ in how much their location and growth patterns will be determined by local capabilities. In some industries, capabilities are only of secondary concern. For instance, industries that rely on natural resources can only locate where those resources are found: fishing requires access to bodies of water and, regardless of which capabilities a city has, mining activities cannot emerge without mineral deposits. Similarly, many nontraded services, such as restaurants and shops, require easy access to large markets, and the presence and size of public services, such as health care and education, are determined by government policies. Many authors therefore restrict their analysis to industries in the private sector that produce tradable products and that are not based on natural resources.

But how relevant are these restrictions? To answer this question, we reanalyze our specification grid for four (mutually exclusive) sectors: public-sector industries, resource-based industries, non-traded services, and all remaining industries, to which we will refer as \emph{traded} industries. The exact definition of each sector is provided in Appendix \ref{app:ind_groups}.

Figs \ref{fig:sectorspecific_levels} and \ref{fig:sectorspecific_growth} show scatter plots of the out-of-sample performance in each sector against the out-of-sample performance in the overall economy. Predictive performance is highly correlated across sectors: specifications that perform well in the overall sample also tend to do so within each sector. This consistency is reassuring: it suggests that our analysis is robust, yielding similar preferred specifications in different subsamples. 

Furthermore, whenever observations lie above the 45 degree line, specifications perform better in the corresponding sector than in the overall economy. This often happens in  the traded sector, but rarely in nontraded services. However, in all four sectors, the principle of relatedness has at least some predictive validity. Moreover, predictive performance is not particularly poor in public sector activities or resource-based industries, suggesting that restricting the sample to traded industries may be helpful, but not necessary. 

There are plausible explanations for why predictability is high also outside non-resource based, private-sector, traded industries. For instance, different resource-based activities may be attracted by the same geological conditions or their presence may be betrayed by the presence of downstream industries. Similarly government services will locate in predictable locations such as regional capitals or following central place theory \citep{christaller1933zentralen}. However, such explanations align poorly with the leading explanation for the principle of relatedness, namely, that the reason why we observe related diversification is that related industries share similar capability requirements. The principle of relatedness would instead reflect also other forces than capabilities. Alternatively, we could stretch the concept of capabilities beyond its common meaning. However, this would make the term less discerning: we typically think of a capability as something that a city can develop through investments and learning, not as something that is predetermined by its location.

\begin{figure}[t!]
	\begin{center}
        \begin{subfigure}[b]{1\textwidth}
        	\centering
        	\includegraphics[scale=.28]{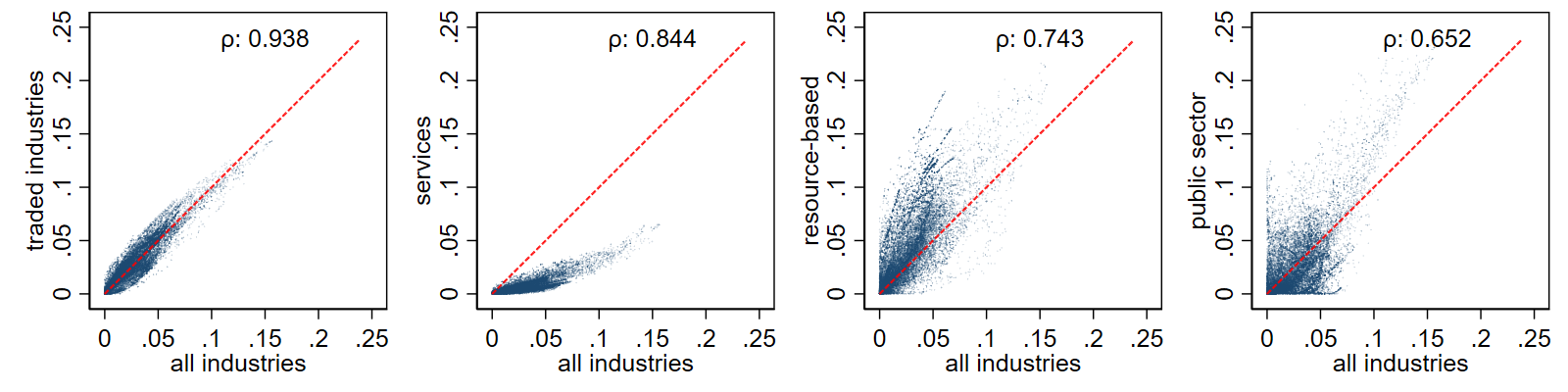}
            \caption{employment levels}\label{fig:sectorspecific_levels}
            \vspace{.5cm}
        \end{subfigure}
        \begin{subfigure}[b]{1\textwidth}
        	\centering
            \includegraphics[scale=.28]{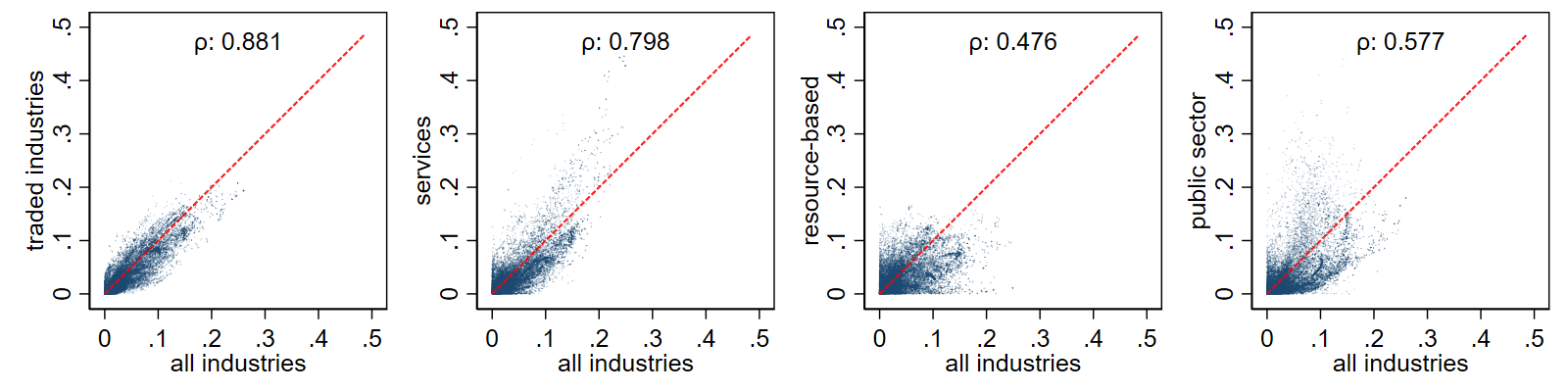}
            \caption{employment growth}\label{fig:sectorspecific_growth}
    \end{subfigure}
    \mycaption{Out-of-sample performance by sector.}{Plots show $\frac{R^2_s - R^2_b}{R^2_b}$, with $R^2_s$ the out-of-sample $R^2$ of specification $s$ and $R^2_b$ of the baseline specification. Each observation in the scatterplots represents a specification in our specification grid, with its performance in the overall sample on the horizontal axis and in a specific subsample on the vertical axis. Panel \ref{fig:sectorspecific_levels} refers to performance in predicting employment levels, panel \ref{fig:sectorspecific_growth} in employment growth.}    \label{fig:sectorspecific}
    \end{center}
\end{figure}

\subsection{Related variety or relatedness?}
 A particularly important specification choice turns out to be whether to base density on continuous information on industries' prevalence in a city as opposed to binary information on their presence. This finding is more than a mere technicality. It has important implications for our understanding of the principle of relatedness. To see this, note that, if in eq. (\ref{eq:density}) we use continuous prevalence information, density quantifies the \emph{size} or \emph{mass} of related industries in the city, reminiscent of Marshallian externalities \citep{rosenthal2004evidence}. In contrast, if we use binarized (presence) information, density becomes a proximity-weighted \emph{count} of related industries. This emphasizes the diversity of activities in a city, reminiscent of Jacobs externalities and related variety benefits \citep{frenken2007related}.

Now, does the principle of relatedness reflect benefits of related variety or of related mass? To analyze this, we construct two versions of our density measures that only differ in whether they use continuous prevalence or binarized presence information. Because performance also depends on the relatedness matrix used, we run separate regressions for each type of productive unit used to calculate relatedness. 

\begin{table}[t!]
    \begin{center}
        \begin{tabular}{l*{5}{c}}
\hline\hline
            &\multicolumn{1}{c}{Establishment}&\multicolumn{1}{c}{Firm}&\multicolumn{1}{c}{City}&\multicolumn{1}{c}{Country}\\
\hline
\multicolumn{5}{c}{\textit{Employment levels}} \\
\hline
$\text{density}_{mass}$    &    0.574***  &      0.533***   &     0.650***   &    0.518***\\
               & (0.0278)     &   (0.0332)      &  (0.0469)      & (0.0382)   \\
$\text{density}_{var}$     &  -0.0863*    &     -0.171***   &    -0.105      &  -0.0518   \\
               & (0.0382)     &   (0.0501)      &  (0.0633)      & (0.0425)   \\
\hline
\(N\)                &      384705         &      384705         &      384705         &      384705   \\
\(R^2\)              &     0.721    &       0.711    &       0.724     &      0.706   \\
\hline
\multicolumn{5}{c}{\textit{Employment growth}}\\
\hline
$\text{density}_{mass}$   &   0.00748***   &   0.00893***  &    0.0104***   &   0.00301   \\
                          &   (0.00121)    &   (0.00130)   &   (0.00172)    &   (0.00154) \\  
$\text{density}_{mass}$   &    0.00371*    &   -0.00109    &   0.000989     &    0.00391  \\ 
                          &    (0.00188)   &    (0.00168)  &    (0.00218)   &    (0.00201)\\ 
\hline
\(N\)                &      245395         &      245395             &      245395         &      245395 \\
\(R^2\)              &      0.068       &    0.068        &   0.072    &      0.065     \\
\hline\hline
\end{tabular}

    \end{center}
  \mycaption{Related variety versus mass of related activity.}{Standard errors in parentheses, \sym{*}: \(p<0.05\), \sym{**}: \(p<0.01\), \sym{***}: \(p<0.001\). Upper panel: regression analysis of employment levels, controlling for industry and city size effects. Lower panel: regression analysis of employment growth, controlling for mean reversion, industry and city size effects. The coefficients reflect the effects of density based on relatedness measured in the productive units listed in the columns, where $\text{density}_{mass}$ uses continuous information on an industry's prevalence in a city, and $\text{density}_{var}$ binarized information on whether or not an industry is significantly present in the city. For results based on our preferred continuous specifications, see Table \ref{tab:variety_vs_mass_con} in Appendix \ref{app:denspairs}.}\label{tab:variety_vs_mass_bin}
\end{table}

Results in Table \ref{tab:variety_vs_mass_bin} offer a surprisingly clear verdict. Regardless of whether we predict employment levels or growth rates, only the density that uses continuous information on the prevalence of an industry in a city is positively and significantly associated with employment and employment growth.\footnote{Table \ref{tab:variety_vs_mass_con} of Appendix \ref{app:denspairs} corroborates this finding, using our preferred continuous specification, where ``continuous'' refers to the way in which we calculate  relatedness, not density.} This finding may explain why the positive effects of related variety on economic development in \cite{frenken2007related} have sometimes proved hard to replicate. For instance, seven out of thirteen studies reviewed by \cite{content2016related} report mixed support for the related variety hypothesis. Although a full analysis of this is beyond the scope of this paper, our analysis suggests that previously reported related variety effect may, in fact, be due to omitted-variable bias: because related variety studies don't control for related mass effects, they may erroneously have attributed the effect of relatedness to related variety.\footnote{Note that related variety studies typically estimate effects at the aggregate level of cities, not industries within cities. Controlling for relatedness would therefore require measures of industrial coherence as in \cite{teece1994understanding}.} 

\section{Policy implications}

Because the principle of relatedness helps identify promising diversification paths, it has started to play an important role in regional development policy frameworks \citep[e.g., ][]{balland2019smart,rigby2022eu,boschma2022evolutionary,hidalgo2022policy}, with the European Union’s Smart Specialization policy as a leading example. These frameworks often use industry space networks to help policy makers prioritize certain industries over others. For instance, \cite{balland2019smart} argue that relatedness proxies how costly it would be for a region to develop a specific industry. They submit that, because related industries (presumably) share many capabilities, moving into industries that are closely related to a region's existing industries economizes on the number of new capabilities the region needs to develop. 

Our results in section \ref{sec:results} support such an interpretation. However, just because it is easy to develop an industry does not mean that it is attractive to do so. Therefore, most authors complement information about inter-industry relatedness with information that captures how attractive an industry is. For this purpose, different alternatives have been proposed. For instance, \cite{balland2019smart} and \cite{rigby2022eu} suggest that regions should try to move into nearby industries that are \emph{complex}. Complex industries require many capabilities and are therefore somewhat shielded from competition from other regions. By developing closely related complex industries, regions can  economize on capabilities, while benefiting from the barriers to entry that characterize these industries. 

However, this approach has several drawbacks. First, because most capabilities are hard to observe, determining the complexity of an industry is not straightforward. \cite{balland2019smart} and \cite{rigby2022eu} rely on the so-called economic complexity index \citep[ECI,][]{hidalgo2009building}, which infers complexity from which cities host which industries. Under certain assumptions, the ECI ranks industries by the technological sophistication they require \citep{schetter2019structural,yildirim2021sorting}. However, \cite{mealy2019interpreting} show that, more generally, the ECI can be interpreted as the best one-dimensional representation of a ``location space'', a network that links locations that host related industries. This means that, by construction, low complexity regions will be close to low complexity industries, and high complexity regions to high complexity industries. As a consequence, the two axes on which the policy framework in \cite{balland2019smart} is based -- relatedness and complexity -- are not independent. Finally, \cite{mcnerney2021bridging} show that the ECI can also be regarded as a city's position along a long-run direction of transformation that is consistent with the short-run diversification patterns described by the principle of relatedness. This debate shows that the interpretation of complexity as a measure of desirability needs to be qualified. 

One solution is to not overly rely on the ECI when determining the desirability of industries. For instance, \cite{hidalgo2022policy} argues that the desirability of an industry depends on policy priorities. Although these priorities may include productivity and employment growth -- which are related to complexity -- they may also reflect other objectives, such as reducing inequality or a region’s carbon footprint. Furthermore, \cite{hausmann2009policies} propose combining the ECI with information on an economy’s productivity to determine whether this productivity fully reflects the complexity of the economy's industrial base. If productivity falls short of what would be expected based on the economy's ECI, there should be scope for raising productivity without diversifying into new activities. In this case, policy should focus on increasing the efficiency with which capabilities are deployed. Only once productivity exceeds the levels implied by the city's industry mix does further diversification become necessary.

Another solution is to change how the principle of relatedness is used to support policy making. Such an approach would also help overcome three other drawbacks of extant policy frameworks that emerged from our specification search. First, as shown in section \ref{sec:sectorspecific}, the principle of relatedness is unlikely to be exclusively driven by capabilities. Most authors acknowledge this either implicitly or explicitly and list a variety of issues that should be considered in smart specialization policies, such as institutions \citep{grillitsch2016institutions} and entrepreneurship \citep{hausmann2003economic,coffano2014centrality}. Second, as noted in section \ref{sec:results_relatedness} about the lack of interpretability of predictions, even if capabilities were its main driving force, the principle of relatedness still does not provide insights into \emph{which} capabilities hold back a city's development. Third, as shown in Table \ref{tab:specapriori}), the principle of relatedness is much better at predicting how large an industry should be in a city today than how much it will grow or shrink in the future: the out-of-sample $R^2$ in predictions of employment levels is about a factor ten higher than in predictions of employment growth. To further illustrate these points, we examine how the principle of relatedness could be used to prioritize candidate industries in practice. 

\subsection*{Selecting industries for growth spurts} 

To make matters concrete, we assume that policy makers are particularly interested in identifying industries that are likely to undergo growth spurts, where we define growth spurts as industries with  $RCA<0.25$ in a city in 2011 that jump to $RCA>1$ in 2019. To identify likely candidates for such growth spurts, we rely on the principle of relatedness to predict industries' local employment in 2011, using our preferred binary specification, with city-based relatedness. The residual of this regression tells us for each city-industry combination by how much employment exceeds  the predictions of the principle of relatedness. Positive residuals suggest that the industry is ``too big'', negative residuals that it is ``too small''. As a consequence, the smaller (more negative) the residual is, the greater a local industry's presumed growth potential will be.

\begin{figure}[t!]
    \begin{center}
        \includegraphics[scale=.6]{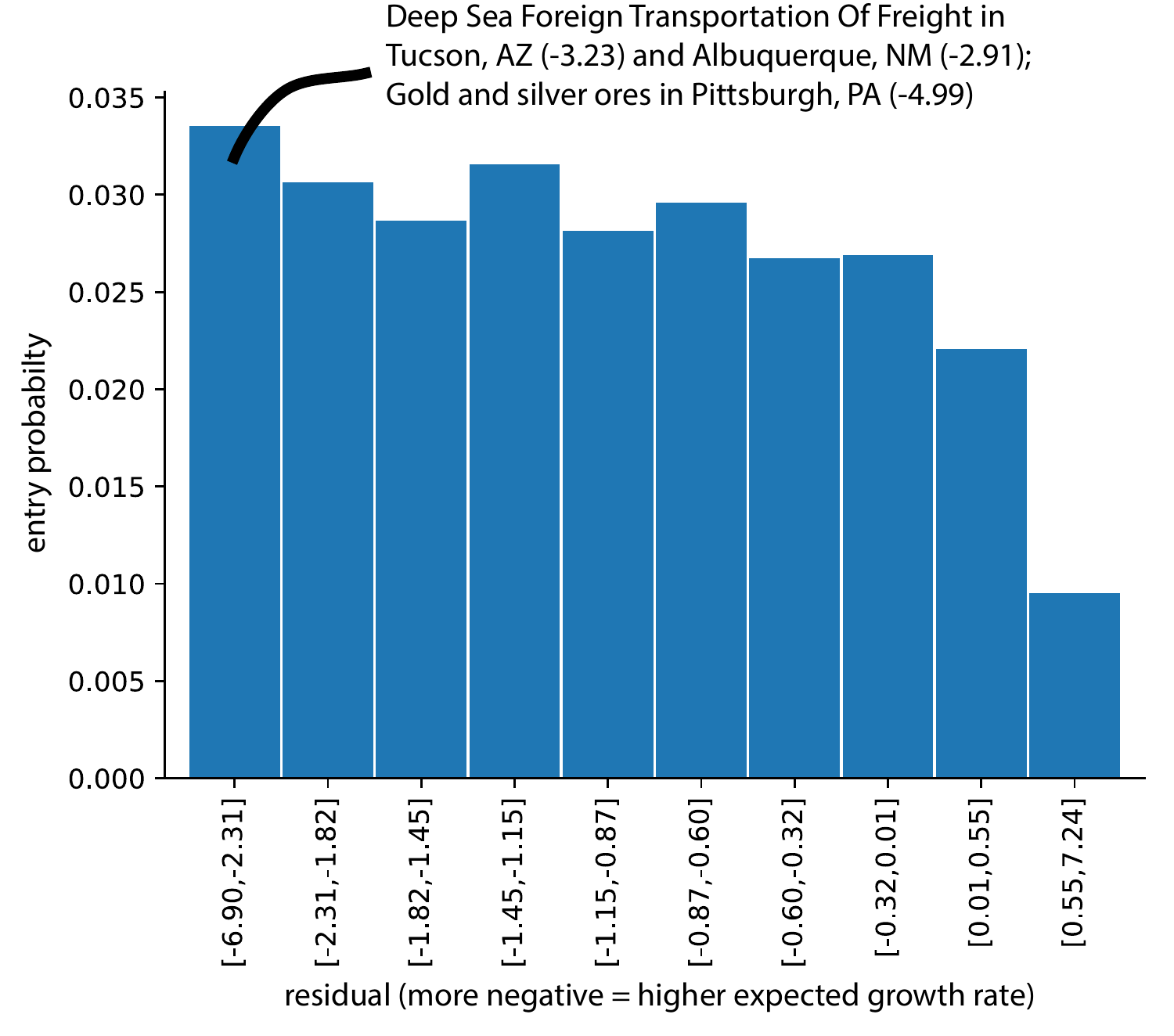}
        \mycaption{Identifying candidates for growth spurts.}{Share of city-industry combinations that undergo a growth spurt. The horizontal axis bins observations based on the residual of a regression that predicts the size of an industry in a city. The more negative this residual is, the more the principle of relatedness would predict the industry to grow.}   \label{fig:policy_figure}
    \end{center}
\end{figure}

 Figure \ref{fig:policy_figure} evaluates how accurate the guesses that we provided to our imaginary policy makers were. The horizontal axis bins all city-industry observations by their estimated residuals. The vertical axis shows which share of local industries in each of these bins undergo growth spurts. 

Results are somewhat mixed. Our advice would certainly have helped policy makers avoid  industries with abysmal growth potential: very few industries with large positive residuals exhibited growth spurts. However, if the goal was to pick winners, Fig. \ref{fig:policy_figure} offers a cautionary tale: even in local industries with very large negative residuals, the likelihood of growth spurts is just marginally above average (3.36\% against 2.68\%). 
 
Why may this have happened? A possible explanation is that the very fact that industries are small in spite of being closely related to the wider local economy  signals that these industry face some (unobserved) hurdles. After all: why, if it were so easy to develop the industry, has the city not done so yet? This suggests leveraging the principle of relatedness for something other than predicting future development paths. Instead, the principle of relatedness may help detect \emph{anomalies} – industries that are active in a region that according to the principle of relatedness shouldn’t be and industries that are not active, while the principle of relatedness predicts they should be.

\subsection*{Anomaly detection and Growth Diagnostics} 
A policy framework that illustrates the value of anomaly detection is Growth Diagnostics \citep[GD,][]{hausmann2008growth}. GD starts from the position that for economic development to happen, many things need to be in place. Applied to industrial development in cities, this means that for an industry to be productive in a city, it needs access to a large number of complementary factors. For instance, to flourish, industries need a highly skilled and diversified workforce, but also functioning infrastructure and utilities, effective institutions, risk-taking entrepreneurs and so on. If one of these factors is in low supply, expanding any of the other factors is unlikely to lead to growth until the initial bottleneck is resolved. To find these bottlenecks, \cite{hausmann2008doing} propose a set of diagnostics tools that aim to identify binding constraints to growth.\footnote{These diagnostic tools try to infer missing factors by observing the behavior of economic actors. In particular, if certain factors are absent, firms and other economic actors will try to find workarounds. Observing these workarounds provides important clues for policy makers about development bottlenecks in their economies.}   


We propose that industry spaces can be a useful addition to this toolbox: although the principle of relatedness may not very accurately identify growth candidates, it reliably pinpoints which industries are surprisingly small or large. Instead of assuming that industries that are too small need to be kick-started by policy prioritization, a more prudent approach would first scrutinize the anomalies that the principle of relatedness detects in a city. 

To illustrate this, we added some examples in Fig. \ref{fig:policy_figure} of city-industry combinations with -- apparently -- high growth potential. These examples were not picked at random. On the contrary, it is quite obvious that Phoenix (AZ) and Albuquerque (NM) are unlikely to develop deep-sea transportation activities, and that opening gold mines in Pittsburgh is probably no promising proposition either. Instead, these high-growth candidates are better understood as anomalies that point to binding constraints to development. In Phoenix and Albuquerque, a prominent constraint is lack of coastal access, in Pittsburgh, lack of gold deposits would explain the city's poor performance in gold mining. 

On the other hand, some industries that are deemed far too large for a city are still unlikely to shrink. For instance, the over 15,000 employees in \emph{Research, development and testing} in Knoxville (TN) represent a very large, positive, anomaly. However, this anomaly is solidly anchored in the city by the presence of the Oak Ridge National Laboratory (ORNL). This research institute has deep roots in the region and is financed by the federal government as part of Department of Energy's system of national research labs. It is therefore unlikely to leave the city. As a consequence, the fact that Research, development and testing is too large according to the principle of relatedness should not be interpreted as a risk for Knoxville, but rather as an opportunity: the city is likely to be able to attract related industries that are synergistic with ORNL's work, some of which are already present, such as \emph{Laboratory apparatus and analytical, optical, measuring, and controlling instruments} and \emph{Engineering, architectural and surveying services}.

\subsection*{The voice of absent industries}
Most anomalies will be harder to explain than the examples listed above. Instead, they require further investigation. The aim of such investigations is to understand which constraints hold back industries' growth in a city. This may involve further statistical analysis, surveys or interviews with private sector representatives. However, the most binding constraints are often those that prevent an industry from locating in a city altogether. Precisely here is where policy makers struggle most to learn about how to lift such constraints: the industries that are absent typically have no ``voice'' in the city, nor do they generate any data (e.g., requests for permits, vacancies, profitability reports) in local administrative records. By highlighting anomalies, the principle of relatedness identifies which absent industries merit further analysis. Policy makers can then invest in targeted analysis, such as approaching firms in these industries  outside the city to ask what prevents them from moving to the policy maker's city.



By focusing less on which activities to support and which not, anomaly detection circumvents the risks associated with picking winners. More importantly, however, using the principle of relatedness as a diagnostic tool can help understand what kind of capabilities are missing in a city. Using several relatedness matrices -- each focusing on a different type of economies of scope -- can help further triangulate these missing capabilities. 

At the same time, the principle of relatedness remains useful in  policy  prioritization. After all, given that the principle of relatedness predicts quite well which industries will \emph{not} exhibit growth spurts, it may still help prioritize (or, better, deprioritize) growth candidates. For instance, the rankings of industries derived from the principle of relatedness are apolitical and objective. This can be used to put a check on the power of well-funded lobbies and vested interests. Anomaly detection should therefore be considered as a refinement, not replacement, of existing policy frameworks.

\section{Discussion and conclusion} \label{sec:conclusion}

The principle of relatedness posits that diversification and growth trajectories of cities are predictable: cities typically grow by developing and expanding industries that are closely related to their current portfolio of industries. We have shown that the principle of relatedness is robust across a wide set of specifications, although there are substantial differences in predictive performance and interpretability across these specifications. Beyond performance differences, different specifications can be seen as testing different ``flavors'' of the principle of relatedness. As a result, our specification search also yields insights into the mechanisms behind the principle of relatedness.

First, different types of productive units combine different industries, resulting in  different types of relatedness. For instance, establishment and firm portfolios exhibit relatedness patterns that conform closer to the sectoral classification system than city and country portfolios. Furthermore, co-occurrences of industries in establishments and firms are strongly shaped by how similar the mix of occupations of different industries is. In comparison, value-chain linkages seem much less important drivers of co-occurrences. Moreover, co-occurrence patterns in cities and countries follow neither human capital nor value chain links very closely. As a consequence, industry spaces -- and the industry clusters derived from them -- are typically more intuitive and readily interpretable when based on establishment or firm portfolios compared to city or country portfolios. At the same time, both city-level and firm-level relatedness matrices perform well in predicting a city's growth trajectory. This suggests that the best compromise between interpretability and predictiveness is provided by firm-based relatedness matrices.

Second, when it comes to measuring density, we show that measures that are based on the \emph{mass} of related industries in a city outperform measures that are based on the \emph{variety} of related industries. This suggests that the principle of relatedness is closer related to Marshallian (i.e., specialization) than Jacobs (i.e., related variety) externalities. 

Third, the leading explanation for the principle of relatedness is that, by moving along trajectories of related diversification, cities can limit the number of new capabilities they need to acquire. However, this raises a puzzle: why is the principle of relatedness also predictive in sectors where capabilities play no primary role in location decisions, such as public services, resource-based industries and nontraded services? A capability-based explanation of the principle of relatedness can only be salvaged if we were to expand the notion of capability far beyond its common meaning. For instance, capabilities would have to include aspects that are hard to change, such as a city's geological conditions or population size. Related to this, we find that the principle of relatedness does not always provide plausible predictions of which industries will grow in a city. In fact, the principle of relatedness seems better suited to identify industries that are least likely to grow than those that are most likely to do so. 

Although none of this invalidates the principle of relatedness, it suggests that relatedness patterns should be interpreted with caution: inter-industry relatedness does not necessarily reflect  shared capabilities. Indeed, understanding \emph{why} industries colocate is an important emerging area of research \citep[e.g., ][]{ellison2010causes, diodato2018industries}.  In light of this, using the principle of relatedness to prioritize industries in industrial policy may be useful, but not without risks. A more conservative approach uses the principle of relatedness to detect anomalies in a city's industrial portfolio. In some cases, these anomalies will be due to missing capabilities, in other cases, they may point to other reasons that hold back a city's development. In either case, these anomalies reveal binding constraints to growth. We have proposed that such anomaly detection is best used as a component of a broader diagnostics approach. Such an approach would complement existing frameworks that help policy makers navigate the trade-off between feasibility and desirability of future development trajectories by focusing on which capabilities a city is lacking and what kind of public policy would help overcome existing development bottlenecks.

\subsection*{Limitations and future research}
Some limitations of our study stand out. First, we have focused on the principle of relatedness in economic geography. That is, we studied its performance in predicting the presence and growth of industries in cities. This choice may explain the exceptional performance of city-level colocation-based relatedness matrices. Future research could analyze specifications and predictive performance of the principle of relatedness applied to growth and diversification in establishments, firms and countries.

Second, in spite of covering many important elements, specifications can be improved in other ways. For instance, we did not consider set-based distance metrics, such as the Jaccard distance. Similarly, when residualizing industry-unit size data, we only considered industry and unit size, leaving aside other characteristics, such as diversity or productivity. Finally, to keep  our estimation framework computationally feasible and allow for millions of repeated analyses, we only explored OLS regressions. Future research could analyze the statistical implementation of our model specifications. 

Third, our results rely on a single database that was not originally intended to offer a representative depiction of the world economy, although it reliably describes the US economy. Further analysis in countries other than the U.S., using, for instance, administrative datasets -- many of which record establishments, firms, industries and locations -- could help corroborate or falsify the findings in this study.

Finally, our analysis focused on industrial development and growth in cities. However, it can be  readily extended to technological and scientific development using patent data or scientific publication data. In both datasets, co-occurrences of technologies or academic fields can be studied at different levels of aggregation: from patents or journal articles to individuals, organizations, cities and countries. It would be particularly interesting to learn how these spheres interact by studying co-occurrences across domains. This would shed light on how diversification dynamics interact across science, technology and the economy.

\bibliography{bibliography}

\newpage
\appendix

\section{Representivity of industry-region aggregates} \label{app:representivity}
\setcounter{table}{0} 
\numberwithin{table}{section}
\setcounter{figure}{0} 
\numberwithin{figure}{section}

\numberwithin{equation}{section}

\subsection{Country aggregates}\label{sub:country_compare}
We use the 2011 wave of the D\&B database to construct relatedness measures at the country level.  To do so, we aggregate the number of employees to 3-digit SIC industries for all 235 countries presented in the database. Next, we select countries for which the D\&B data offer a reasonably accurate depiction of the industrial composition of the national economy.

We compared the industrial employment composition for each country in the D\&B data to data provided by the International Labor Organization, which provides estimates of country-level employment in 14 aggregated sectors based on ISIC rev.4 classification  (the table of employment by sex and economic activity, ILO\// EMP\_2EMP\_SEX\_ECO\_NB). After harmonizing industry classifications, we calculate Pearson correlation coefficients,  for industrial employment shares in both data sources. We furthermore calculate these correlations after excluding the agriculture sector, which may be dominated by self-employed workers, especially in developing economies and therefore be poorly covered by the D\&B data.

Next, we retain countries if:
\begin{enumerate}
    \item the country has at least 100,000 employees in the D\&B data,
    \item the correlation of employment structure is at least 0.4 for all sectors, or at least 0.5 after excluding the agriculture sector,
\end{enumerate}
This selection results in the following list of included countries:
\begin{quote}
AGO, ARE, ARG, AUS, AUT, BEL, BGD, BGR, BHR, BIH, BLR, BOL, BRA, CAN, CHE, CHL, CHN, CMR, COL, CRI, CYP, CZE, DEU, DNK, DOM, ECU, ESP, EST, ETH, FIN, FRA, GBR, GRC, GTM, HKG, HND, HRV, HTI, HUN, IDN, IND, IRN, ISL, ISR, ITA, JAM, JOR, JPN, KAZ, KHM, KOR, LBN, LKA, LTU, LUX, LVA, MAR, MDG, MEX, MKD, MLT, MMR, MOZ, MUS, MYS, NIC, NLD, NOR, NPL, NZL, PAK, PAN, PER, PHL, POL, PRT, PRY, QAT, ROU, RUS, SAU, SDN, SGP, SLV, SVK, SVN, SWE, SYR, THA, TUN, TUR, TWN, UKR, URY, USA, VEN, VNM, YEM, ZAF, ZMB
\end{quote}

Note that data from these countries is only used to calculate country-level relatedness matrices. All other analyses are run on the US segment of the D\&B data only.

\subsection{US aggregates}\label{sub:us_compare}

We compare the 2011 D\&B data with 2011 US County Business Patterns (CBP) data for the analysis. Since CBP is released in NAICS classification but D\&B 2011 in SIC codes, we harmonize the classifications and aggregate to the city-industry level to check the representativity of total employment and employment shares. Results are reported below.

\begin{figure} 
	\begin{center}
	\includegraphics[scale=.55]{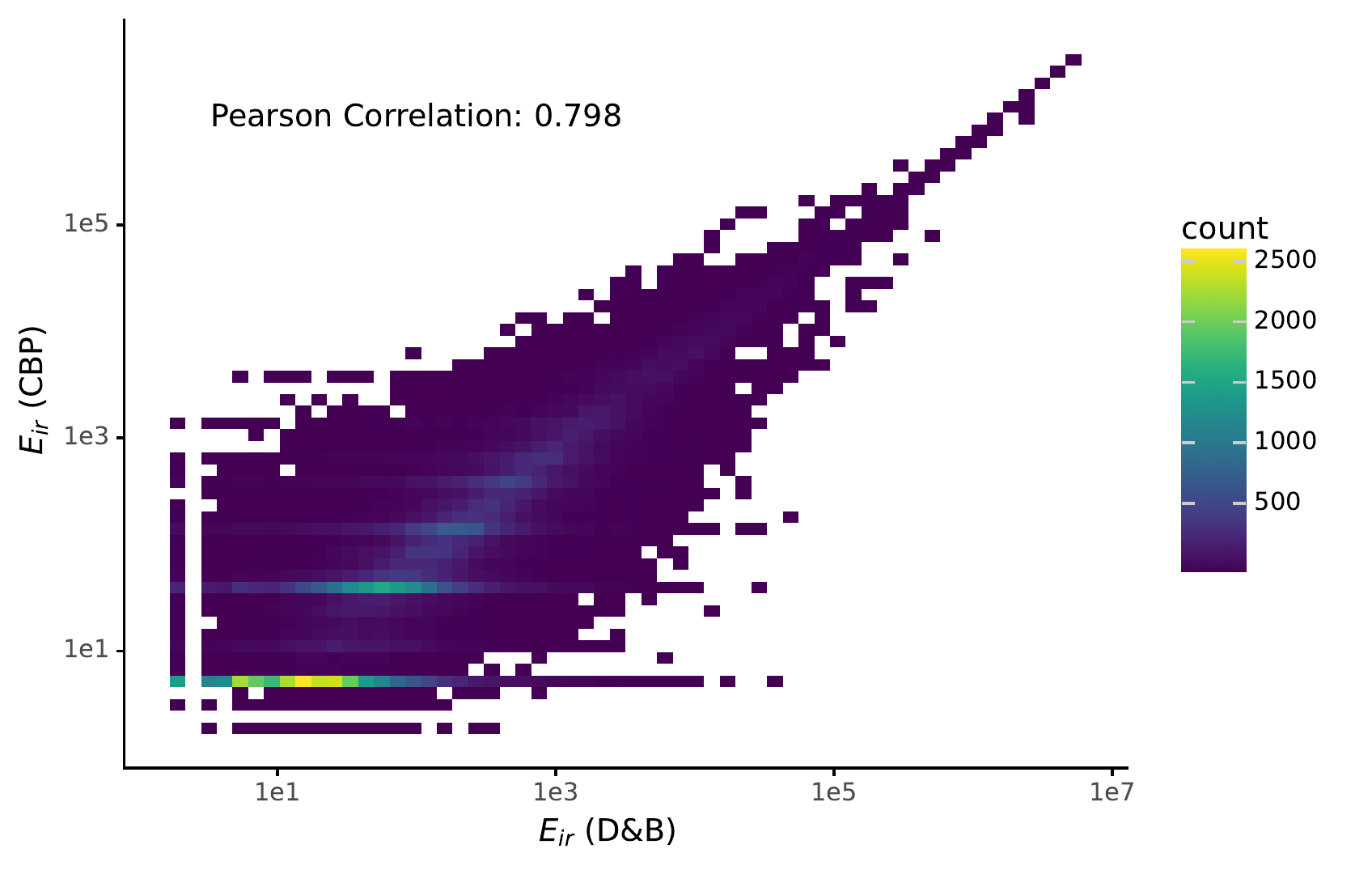}
    \caption{Comparison of employment in US city-industry cells $\left( E_{ir} \right)$: CBP against D\&B}\label{fig:cbpfig1}
	\end{center}
\end{figure}

\begin{figure} 
	\begin{center}
	\includegraphics[scale=.55]{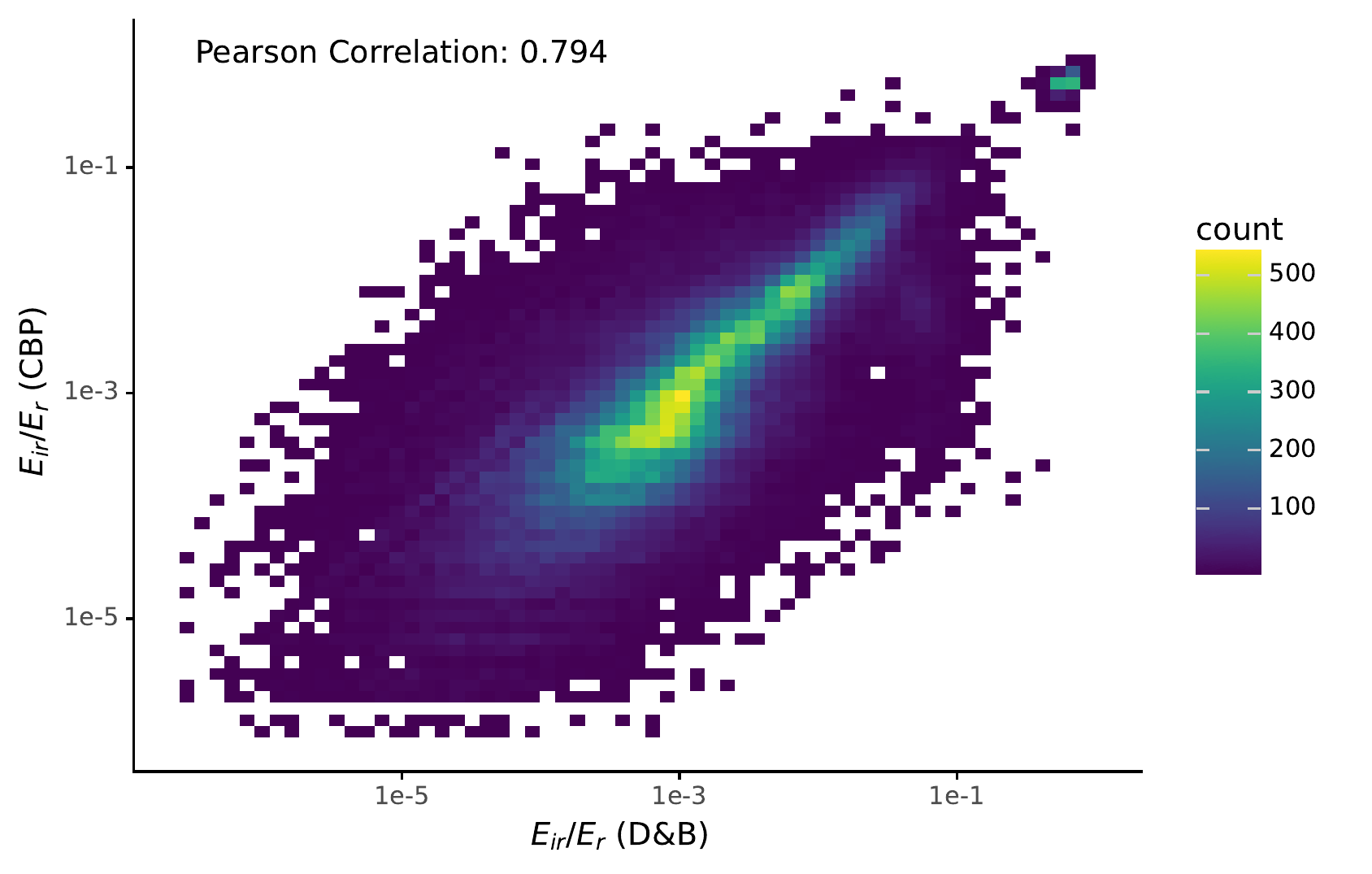}
    \caption{Comparison of industry employment shares in US cities $\left( E_{ir}/E_{r} \right)$: CBP against D\&B
}\label{fig:cbpfig2}
	\end{center}
\end{figure}

\newpage
\section{Setting up the specification grid (Fig. \ref{fig:grid})} \label{app:grid}

\subsection{Prevalence}\label{sub:prevalence}
The simplest way to quantify prevalence is by the \emph{size} of the industry in the productive unit. In this paper, we measure size in terms of numbers of employees, possibly log-transformed to reduce distributional skew. For firms, cities and countries, we only use the primary industry codes of establishments. Next, we attribute all employment in that establishment to its primary industry when aggregating to higher level units. This avoids that all units mechanically pick up the same co-occurrences that are already present at the establishment level, allowing us to more clearly delineate different relatedness types. For establishments themselves, we impute the share of employees corresponding to each industry code, as described in the Appendix \ref{app:est_emp_share}.

A downside of measuring prevalence in terms of raw size is that it doesn't account for the fact that industries differ vastly in size, as do productive units. As a consequence, large industries and large productive units tend to dominate industry co-occurrences. To counteract this, prevalence is often expressed in relation to a benchmark. For instance, \cite{hidalgo2007product} use revealed comparative advantage\footnote{In economic geography, this index is better known as the location quotient.} \citep[RCA, see ][]{balassa1965trade} to determine which product categories have a significant presence in a country's export basket. To be precise, the RCA benchmarks an industry's relative size in a productive unit against its relative size in the overall economy:
\begin{equation}\label{eq:RCA}
    \nu_{iu}^{RCA} = RCA_{iu} = \frac{E_{iu}/E_{u}}{E_{i}/E},
\end{equation}
where $E_{iu}$ is industry $i$'s employment in unit $u$ and omitted subscripts indicate summations over the corresponding dimension. To reduce distributional skew, we can map this measure onto the interval $[0,1)$, using the following transformation \citep[see, e.g., ][]{neffke2017inter}:
\begin{equation}\label{eq:RCA_T}
    \nu_{iu}^{RCA^{*}} = RCA^{*}_{iu} = \frac{RCA_{iu}}{RCA_{iu}+1},
\end{equation}

Alternatively, \cite{van2023information} propose taking logarithms:
\begin{equation}\label{eq:PMI}
    \nu_{iu}^{PMI} = \log RCA_{iu} = \log \frac{\frac{E_{iu}}{E}}{\frac{E_{u}}{E} \frac{E_{i}}{E}} = \log \frac{p_{iu}}{p_{i}p_{u}},
\end{equation}
where $p_{iu}$ is the probability that a randomly drawn worker works in industry $i$ and in unit $u$, $p_{i}$ the probability that the worker works in industry $i$ and $p_{u}$ that she works in unit $u$.\footnote{To avoid problems related to values of $\log0$, we increase all employment counts by one unit before calculating RCAs. \cite{van2023information} show that this amounts to  Bayesian estimation of the PMI with a uniform prior.}

Eq. (\ref{eq:PMI}) is known as point-wise mutual information (PMI). It provides a useful information-theoretic interpretation of eq. (\ref{eq:RCA}): the (logarithm of the) RCA quantifies the amount of surprise in observing a local industry of size $E_{iu}$ when the nation-wide size of the industry is $E_i$ and the unit has a total employment of $E_u$. This amount of surprise may offer a more accurate signal of a productive unit's underlying strengths than the size of the industry does. 

However, the PMI's benchmark presupposes that industries scale linearly with the sizes of the productive units in which they are found. The notion of surprise in the PMI may therefore be rather restrictive.\footnote{The urban scaling literature has shown that industries exhibit a range of scaling coefficients.  For instance, small, complex industries often concentrate in large cities, whereas larger, more ubiquitous industries are spread across cities of different sizes \citep[e.g.,][]{gomez2016explaining,youn2016scaling,hong2020universal,balland2020complex}.} More generally, it may be possible to predict the size of an industry in a productive unit from some general characteristics of the unit and the industry. This suggests to define surprise against conditional expectations, \emph{given such characteristics}. One way to form such conditional expectations is regression analysis, as proposed by \cite{neffke2013skill}. We consider three such benchmarks:
\begin{equation}\label{eq:prevalence_regression_OLS}
    \log E_{iu} = \alpha + \beta_{1} \log E_{i} + \beta_{2} \log E_{u} + \nu_{iu}^{OLS},
\end{equation}
\begin{equation}\label{eq:prevalence_regression_FE}
    \log E_{iu} = \delta_{i} + \gamma_{u} + \nu_{iu}^{FE},
\end{equation}
\begin{equation}\label{eq:prevalence_regression_POI}
    P\left[ Y_{iu} = E_{iu}| X {\tilde\beta} \right] = \frac{e^{-\exp\left\{{X_{iu} \tilde{\beta}}\right\}}\exp{\left\{X_{iu} \tilde{\beta}\right\}^{E_{iu}}}}{E_{iu}!}, \nu_{iu}^{POI}:=\log\left( E_{iu}+1\right) - \log\left( \hat{E}_{iu}+1\right)
\end{equation}

Prevalence is now defined as the residual of these regressions. To avoid the problem associated with the logarithm of 0, we increase $E_{iu}$ by 1 before taking logarithms in these equations. Different regression models yield different residuals: eq. (\ref{eq:prevalence_regression_OLS}) uses an OLS regression with the size of the industry and of the unit as explanatory variables; eq. (\ref{eq:prevalence_regression_FE}) instead uses industry and unit fixed effects; and eq. (\ref{eq:prevalence_regression_POI}) a Poisson regression with the same explanatory variables as eq. (\ref{eq:prevalence_regression_OLS}).

Finally, many authors \citep[e.g., ][]{hidalgo2007product,muneepeerakul2013urban,zhu2017jump,balland2019smart,rigby2022eu} binarize prevalence indices to arrive at an index of industrial \emph{presence}: a dichotomous variable that marks whether or not an industry has a significant presence in a productive unit. Although this procedure ignores some information, it suppresses distributional skew and noise in the tails of prevalence indices. To mimic this approach, we binarize prevalence metrics as follows:
\begin{gather*}
\pi_{iu}^{raw} = 1\left( E_{iu}>0 \right), \\
\pi_{iu}^{RCA} = 1\left( RCA_{iu}>1 \right), \\
\pi_{iu}^{OLS} = 1\left( \nu_{iu}^{OLS}>0 \right), \\
\pi_{iu}^{FE} = 1\left( \nu_{iu}^{FE}>0 \right), \\
\pi_{iu}^{POI} = 1\left( \nu_{iu}^{POI}>0 \right).
\end{gather*}
where $1(.)$ is an indicator function that evaluates to 1 if its argument is true. Note that because because RCA* and PMI are monotonous transformation of RCA, they yield the same binarized versions. The same holds for $E_{iu}$ and $\log E_{iu}$

\subsection{Relatedness}\label{sec:relatedness}

We can summarize prevalence information for an entire economy in $(N_u \times N_i)$ prevalence matrices, $V$, that collect industries' prevalence across productive units. In the same way, we can use presence metrics to create $(N_u \times N_i)$ presence matrices, $\Pi$. Outcome-based relatedness indices transform these prevalence or presence matrices into $(N_i \times N_i)$ inter-industry relatedness matrices, $\Phi$. 

Extracting relatedness matrices from prevalence matrices has a long history in the field of scientometrics \citep[e.g.,][]{engelsman_mapping_1991, leydesdorff2006co, leydesdorff2008normalization}. Compared to this literature, We focus on approaches that have been used in research on diversification dynamics of regional economies. In this context, the principle of relatedness furthermore provides us with a clear task in which to evaluate their predictive performance.

One set of methods operates directly on the presence or prevalence matrix. A column of such a matrix records how the employment of the corresponding industry is distributed across productive units. The similarity of different columns can now be regarded as a measure of the relatedness between the corresponding industries. We will consider two such measures. The first is the correlation between pairs of columns $(i,j)$ in a $\Pi$ or $V$ matrix as in \cite{hausmann2021implied}:

\begin{equation}\label{eq:correlation}
    \phi_{ij}^{corr} = 1 + \tfrac{1}{2}corr_{u \in U}(\nu_{ui},\nu_{uj}),
\end{equation}
where $corr_{u \in U}$ expresses the correlation across productive units $u$, which is then rescaled to map onto the interval $[0,1]$, and $U$ the set of all productive units in the economy. Two industries are thus related if they exhibit correlated prevalence patterns across productive units.

As an alternative, we consider the cosine similarity between prevalence vectors:
\begin{equation}\label{eq:cosine_dist}
    \phi_{ij}^{cos} =1 + \tfrac{1}{2} \frac{\sum\limits_{u \in U}v_{iu}v_{ju}}{\sqrt{\sum \limits_{u \in U} \nu_{iu}^{2} \sum \limits_{u \in U} \nu_{ju}^{2}}}.
\end{equation}

A different set of methods first constructs a so-called co-occurrence matrix: 

\begin{equation}\label{eq:cooc_bin}
    c_{ij} = \sum \limits_{u \in U} \pi_{iu} \pi_{ju},
\end{equation}
where $\pi_{iu} = 1(v_{iu}>\xi)$, with $\xi$ some threshold value. Or, in matrix form:
\begin{equation}\label{eq:cooc_bin_mat}
    C = \Pi' \Pi,
\end{equation}
where $C$ is the co-occurrence matrix and $\Pi'$ the transpose of presence matrix $\Pi$.

Co-occurrence matrices count the number of times that two industries are present in the same productive units. Although they are typically based on binarized presence matrices, eq. (\ref{eq:cooc_bin_mat}) immediately suggests variants that use continuous prevalence information:
\begin{equation}\label{eq:cooc_cont_mat}
    \tilde{C} = V' V,
\end{equation}
where $V$ represents the matrix of industry-unit prevalences and $\tilde{C}$ the matrix of co-prevalences.

Co-occurrence matrices are typically normalized to account for the fact that some industries are present in more productive units than others. To do so,  \cite{hidalgo2007product} propose calculating the minimum of two conditional probabilities:
\begin{equation}\label{eq:min_prob}
    \phi_{ij}^{MCP} = \min\left(\frac{c_{ij}}{c_{i}},\frac{c_{ij}}{c_{j}}\right),
\end{equation}
where the first fraction represents the probability of observing industry $j$ in a productive unit, given that we had already observed industry $i$ in the same unit, and the second term the probability of observing $i$ given that we observed $j$ in the unit. 

We can rewrite eq. (\ref{eq:min_prob}) as:
\begin{equation}
    \label{eq:generalized_min_prob_orig}
    \phi_{ij}^{MCP} = \frac{c_{ij}}{\max(c_{i},c_{j})}.
\end{equation}

This shows that eq. (\ref{eq:min_prob}) effectively ignores the size of the smallest industry when creating a benchmark for the observed co-occurrences. This choice can be relaxed in a family of normalizations of the form:
\begin{equation}
    \label{eq:generalized_min_prob}
    \phi_{ij}^{\kappa} = \frac{c_{ij}}{\min(c_{i},c_{j})^\kappa \max (c_{i},c_{j})^{1-\kappa}},
\end{equation}
where $0 \leq \kappa \leq 1$. When $\kappa=0$, eq. (\ref{eq:generalized_min_prob}) reduces to the minimum conditional probability of eq. (\ref{eq:min_prob}), whereas $\kappa=0.5$ puts equal weights on the sizes of both industries.\footnote{Small $\kappa$s avoid high relatedness estimates that are driven by small industries, limiting the risk of spuriously high relatedness estimates.}

Another intuitive normalization is analogous to the RCA transformation of eq. (\ref{eq:RCA}):\footnote{See, for instance, \cite{neffke2017inter}}
\begin{equation}\label{eq:cooc_bin_norm}
    \phi_{ij}^{RCA} = \frac{c_{ij}}{c_{i}c_{j}}c,
\end{equation}

To reduce distributional skew, we can apply the same transformation as in eq. (\ref{eq:RCA_T}),  casting the relatedness value between 0 and 1.

What all of these metrics have in common is that they leverage the information available in the co-prevalence,  $\tilde{c}_{ij}$, or co-occurrence,  $c_{ij}$, terms. In Appendix \ref{app:relation_metrics}, we show that this also holds for the cosine and correlation-based metrics, which turn out to be closely related to eq. (\ref{eq:cooc_cont_mat}).



Finally, some authors \citep[e.g.,][]{muneepeerakul2013urban} argue that \emph{positive relatedness} (i.e., industries that coincide surprisingly often in productive units) is fundamentally different from \emph{negative relatedness} or \emph{unrelatedness} (i.e., industries that are combined surprisingly little). Whereas the former points to the existence of economies of scope or shared capability requirements, the latter indicates that industries generate negative externalities for, or compete with, one another. Therefore, for each relatedness matrix, we also create a version in which we \emph{truncate} relatedness. That is, we set all elements that correspond to negative relatedness to the value that represents neutral relatedness, therewith ignoring negative relatedness.\footnote{In correlation- or cosine-based relatedness matrices, this value is equal to zero, in RCA-based relatedness matrices, the value corresponding to neutral relatedness equals one. Because  negative relatedness is ill-defined for the conditional-probability-based approaches, we do not pursue this truncation strategy there. Moreover, because the positive part of cosine and correlation measures already lies in the interval $[0,1]$ we do not need to rescale and recenter these measures as in eqs (\ref{eq:correlation}) and (\ref{eq:cosine_dist}).}





\subsection{Density}\label{sec:density}
Relatedness matrices are typically rather sparse: whereas a small number of combinations of industries exhibit strong ties, most industries are unrelated or weakly related. In this sense, relatedness matrices describe for each industry a natural \emph{ecosystem} of other industries that should help the industry thrive. By estimating the size of these ecosystems, we can get a sense of how well an industry fits the industrial mix of an entire local economy.

To do so, \cite{hidalgo2007product} introduce a variable that they call \emph{density}. The density of industry $i$ in city $r$, $d_{ir}$, is constructed as the weighted average prevalence of all other industries in city $r$, where the weights express industry $i$'s relatedness to these other industries. Because most industries are weakly related or unrelated, some authors only use the $k$ most related neighbors of $i$ in this calculation \citep[e.g.][]{hausmann2021implied}. Taken together, this yields eq. (\ref{eq:density}), replicated here for convenience:

\begin{equation*}
    d_{ir} = \sum \limits_{j \neq i \in J_{i}} \frac{\phi_{ij}}{\sum \limits_{k \neq i \in J_{i}}\phi_{ik}} \nu_{jr}.
\end{equation*}

\subsection{Relation between co-prevalence and co-occurrence metrics}\label{app:relation_metrics}

The conversion of $(N_u \times N_i)$ industrial prevalence or industrial presence matrices into $(N_i \times N_i)$ inter-industry relatedness matrices relies on rescaling the intensity of co-prevalences or co-presences of industries across productive units. Measures may differ in the way they rescale such intensities, although they share commonalities at a more fundamental level. Here we illustrate this by comparing relatedness measures based on continuous prevalence information to their discrete counterparts. First, note that $\phi_{ij}^{corr}$ in eq. (\ref{eq:correlation}) can be written as:

\begin{equation}\label{eq:correlation_explained}
    \phi_{ij}^{corr} = \frac{\frac{1}{N_{u}}\sum\limits_{u \in U}v_{iu}v_{ju} - \frac{1}{N_{u}} \left( \sum \limits_{u \in U} v_{iu}\right) \frac{1}{N_{u}} \left( \sum \limits_{u \in U} v_{ju}\right)}{\sqrt{\frac{1}{N_{u}}\sum \limits_{u \in U} v_{iu}^{2} - \left(\frac{1}{N_{u}} \sum \limits_{u \in U} v_{iu}\right)^{2}} \sqrt{\frac{1}{N_{u}}\sum \limits_{u \in U} v_{ju}^{2} - \left(\frac{1}{N_{u}}\sum \limits_{u \in U} v_{ju}\right)^{2}}},
\end{equation}
where $N_{u}$ denotes the number of different productive units in the economy.

Also note that $\phi_{ij}^{cos}$ in eq. (\ref{eq:cosine_dist}):
\begin{equation}
    \phi_{ij}^{cos} = \frac{\sum\limits_{u \in U}v_{iu}v_{ju}}{\sqrt{\sum \limits_{u \in U} v_{iu}^{2} \sum \limits_{u \in U} v_{ju}^{2}}},
\end{equation}
is essentially equivalent to $\phi_{ij}^{corr}$ in eq. (\ref{eq:correlation_explained}), except for the fact that $\phi_{ij}^{cos}$ does not mean-center prevalences. 

Now compare these continuous co-prevalence metrics to the following expression for the binary co-occurrence approach of $\phi_{ij}^{RCA}$ in eq. (\ref{eq:cooc_bin_norm}):

\begin{equation}
    \phi_{ij}^{RCA} = \frac{\sum \limits_{u \in U} v_{iu} v_{ju}}{\sum \limits_{u \in U, k \in J} v_{iu} v_{ku} \sum \limits_{u \in U, l \in J} v_{ju} v_{lu}}\sum \limits_{u \in U, l \in J, m \in J} v_{lu} v_{mu},
\end{equation}
where $J$ represents the set of all industries in the economy.

The two expressions are very similar. Like the correlation approach, the main signal of relatedness is derived from the term $\sum \limits_{u \in U} v_{iu} v_{ju}$, which expresses the co-prevalence or co-occurrence of two industries $i$ and $j$. However, the cosine metric normalizes this co-prevalence by the raw second moments of the prevalence distributions of $i$ and $j$, whereas the correlation metric normalizes by their centered second moments, i.e., their variances. 

The $\phi_{ij}^{RCA}$ metric, in contrast, normalizes  $\sum \limits_{u \in U} v_{iu} v_{ju}$ by the frequency with which industry $i$ and $j$ participate in co-occurrences.\footnote{Note that the final term, $\sum \limits_{u \in U, l \in J, m \in J} v_{lu} v_{mu}$, just rescales the metric for all industry pairs to ensure that a neutral level of co-occurrence coincides with a value of 1.} Therefore, it penalizes industries more heavily that tend to find themselves in large productive units, i.e., in productive units that host many industries. 

In fact, if we also feed dichotomous data into eq. (\ref{eq:cosine_dist}) and if each productive unit holds exactly two industries -- such that the number of co-occurrences in which an industry participates equals the number of productive units that host it -- we arrive at almost the same rescaling factor as in eq. (\ref{eq:cooc_bin_norm}). To see this, note that we can write the denominator of eq. (\ref{eq:cooc_bin_norm}) as:
\begin{equation}
    \sum \limits_{u \in U, k \in J} v_{iu} v_{ku} \sum \limits_{u \in U, l \in J} v_{ju} v_{lu} = S_i S_j,
\end{equation}
where $S_a$ denotes the number of productive units in which industry $i$ is found, and of eq. (\ref{eq:cosine_dist}) as:
\begin{equation}
    \sqrt{\sum \limits_{u \in U} v_{iu}^{2} \sum \limits_{u \in U} v_{ju}^{2}} = \sqrt{S_i S_j}.
\end{equation}

Therefore, differences between relatedness metrics reflect how we penalize large industries. Do we want to penalize industries with a large overall presence or prevalence, as in the correlation or cosine metric? This penalization can then be done in a scale-invariant way as in the correlation metric, which mean-centers the correction terms, or not, as in the cosine metric. Or do we want to penalize industries with large co-prevalences or many co-occurrences as in the $\phi_{ij}^{RCA}$? In this case, we account for the fact that some industries tend to be overrepresented in large productive units, whereas others are overrepresented in smaller productive units. Finally, we can choose the penalization to be asymmetric as in $\phi_{ij}^{\kappa}$.

\newpage
\section{Results and detailed recommendations} \label{app:results_and_recommendations}

Below, we describe a number of general conclusions that can be drawn from Tables \ref{tab:relatedness} and \ref{tab:density}. Based on these conclusions, we also propose a number of recommendations about how \emph{not} to approach the principle of relatedness.

\subsection{Determinants of performance}\label{app:Shapley_value}
\begin{figure}[h!]
	\begin{center}
	\includegraphics[scale=.5]{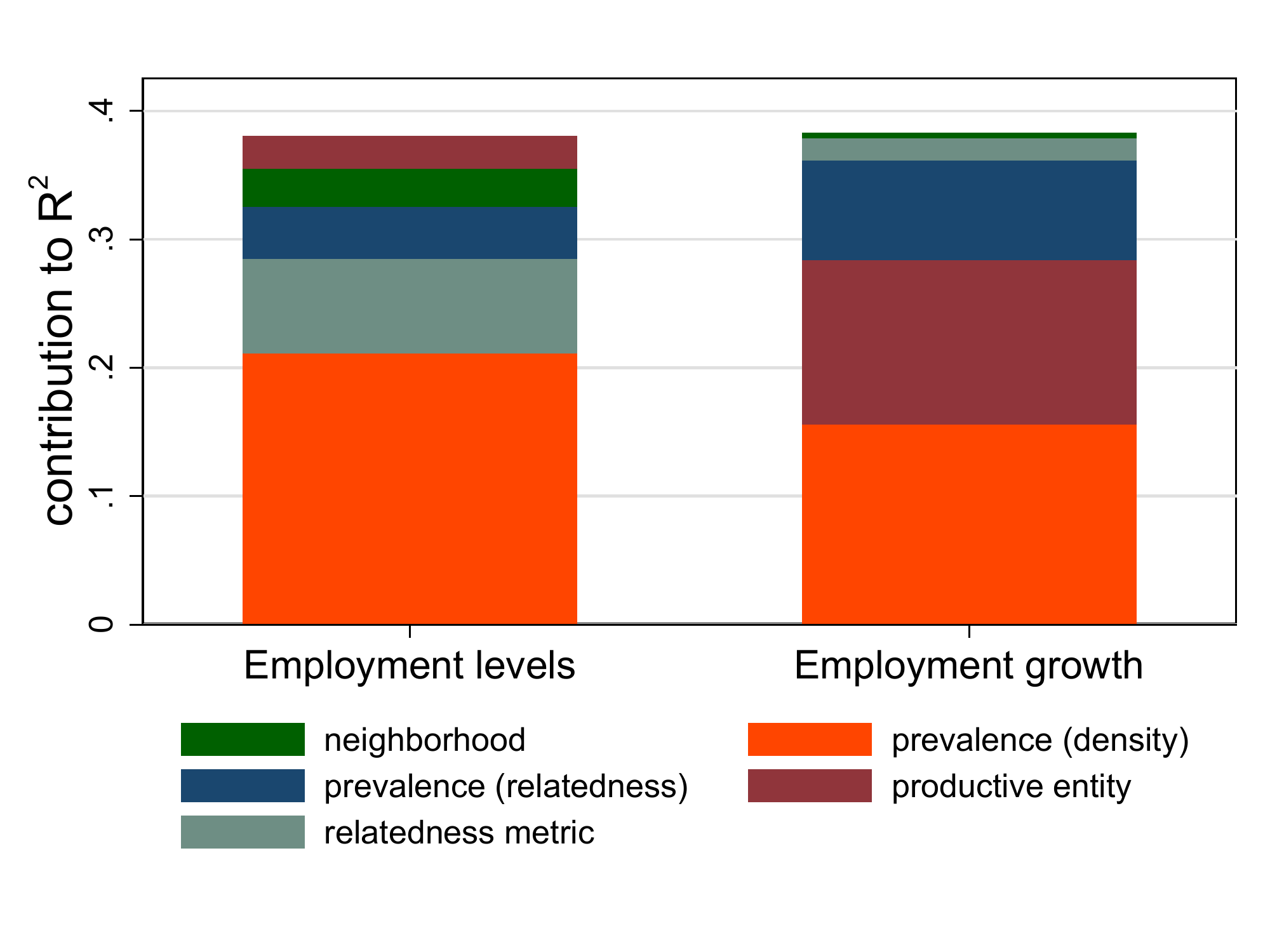}
    \mycaption{Dominance analysis.}{Contribution of specification characteristics to the $R^2$ of OLS regressions that explain model performance from specification characteristics.}    \label{fig:dominance}
	\end{center}
\end{figure}

To determine which aspects of the specification grid have the largest impact on predictive performance, we regress predictive performance on specification characteristics. Fig. \ref{fig:dominance} shows how much each aspect contributes to the $R^2$ of this regression. To determine these contributions, we attribute fractions of the $R^2$ of a model that tries to explain performance from specification characteristics to the different specification elements  using dominance analysis. This procedure is an application of the Shapley value: it generates all $2^k-1$ possible combinations of regressors and estimates the model fit for each combination. Next, for each variable, it asks by how much the average model fit drops if the variable is removed from all covariate sets of which it is a part. In this procedure, we code the main axes of our specification grid as dummy groups that are included or excluded as sets and use as a dependent variable the performance in employment or employment growth predictions.

The most relevant aspect of the model specification is how we define density and, in particular, how we define the prevalence of an industry in a city in the construction of this variable. Other specification characteristics matter less, although for predicting growth, the choice of the productive unit also has a large impact. The least important aspect seems to be how many neighbors we consider when constructing density variables. 

Below, we analyze in greater detail how different specification elements affect the robustness of a specification's performance.

\paragraph{Productive unit.} Regardless of binarization, the city level seems to be the most robust choice for relatedness calculations. The next best choice is the firm level. Country and establishment portfolios, in contrast, hardly ever result in highly robust specifications.

Interestingly, the performance advantage of city-based relatedness matrices is driven by their capacity to predict employment \emph{levels}. If we only consider employment growth, most of the best-performing specifications rely on firm-based relatedness matrices: of all specifications that rely on firm-based relatedness measures, 19.4\% rank in the top decile when it comes to predicting growth patterns. For specifications relying on city-based relatedness measures this share is  10.4\%,  comparable to the 10.2\% for specifications relying on establishment-based relatedness measures, but much better than the 0.1\% for country-based relatedness measures.

It is furthermore important to note that the city and firm levels give rise to qualitatively different industry-unit prevalence matrices. With over 80,000 firms and only 3.14 industries per firm (see Table \ref{tab:summarystat}), the firm-prevalence matrix is very sparse, featuring fewer than 1\% non-zero elements. In comparison, with only 927 cities, over two thirds of the city-prevalence matrix consist of nonzero elements. In section \ref{sec:results_relatedness}, we show that these differences also lead to qualitative differences in the topology of the relatedness networks encoded in the relatedness matrices. In particular, firm-based matrices seem more ``structured'' than their city-based counterparts. This may explain, why truncating relatedness matrices and limiting neighborhoods is particularly important when using city-based relatedness matrices -- which feature many noisy elements -- but less so when using firm- or  establishment-based matrices, in which elements are either zero or very large, leaving little doubt about whether or not two industries are related. Similarly, the implicit recommendation of Table \ref{tab:relatedness} of using small values of $\kappa$ reduces noisy links created by small industries. This recommendation is more forceful for city-based than for firm-based measures.

\paragraph{Proximity metric.} Regardless of the productive unit, the general approach of \cite{hidalgo2007product} to calculating relatedness works remarkably well. \citeauthor{hidalgo2007product} rely on a binarized co-occurrence approach according to which an industry is present in a unit when its $RCA>1$ and relatedness is defined using minimum conditional probabilities. This generalizes to the less restrictive versions of this principle introduced in eq. (\ref{eq:generalized_min_prob_orig}), as long as  $\kappa$ is well below 0.5 to ensure that most emphasis is put on the probability that conditions on the presence of the larger of the two industries in each pair. In Appendix \ref{APP:gridfigs}, we analyze the choice of $\kappa$ more extensively. In general, $\kappa$s should be kept relatively close to 0. However, when predicting growth rates, higher values, of around 0.2 or 0.3, also often yield high performance.  Only when relatedness is based on urban colocation patterns, defining proximity as the RCA of co-occurrences may perform even better. If, instead, continuous coprevalence information is used, matters are less clear-cut. For city coprevalences, using RCAs, provided that they are suitably transformed to reduce skew (as in RCA* or PMI), suffices. However, in firm portfolios, an industry's prevalence is best determined as the residual from the OLS regression of eq. (\ref{eq:prevalence_regression_OLS}). This procedure  yields good results across productive units. Moreover, further experiments (not shown) that use counties instead of cities as a regional unit also suggest that OLS regressions generate the most useful benchmark to determine an industry's prevalence in a productive unit.

\paragraph{Density.} The aspect that has the largest impact on a specification's performance is how we assess the prevalence of an industry in a city when calculating densities. Table \ref{tab:density} shows that robustness is highest when density variables are based on continuous prevalence measures. In particular, comparing an industry's size in a city to its predicted size based on the OLS models of eq. (\ref{eq:prevalence_regression_OLS}) works well. Furthermore, it is helpful to focus on positive relatedness and neglect negative relatedness, supporting the argumentation in \cite{muneepeerakul2013urban}. Finally, the number of closely-related ``neighboring'' industries is not terribly important, but ideally kept small, at between 10 and 20\% of the total number of industries in the economy.

\subsection{Recommendations}\label{app:recommendations}
The findings in Tables \ref{tab:relatedness} and \ref{tab:density} do not yield a single best specification for quantifying the principle of relatedness. However, we can identify a number of specification choices that should be avoided, because they generally perform poorly:

\emph{When constructing inter-industry relatedness matrices}

\begin{itemize}
    \item don't define an industry's prevalence in a productive unit in terms of its raw size, but evaluate this size against a benchmark;
    \item don't use complicated models (i.e., with unit and industry fixed effects or Poisson models) to generate these benchmarks: simple OLS specifications typically work well;
    \item don't use prevalence variables with highly skewed distributions such as the RCA, but instead transform these variables using, for instance, eq. (\ref{eq:RCA_T});
    \item don't use values of $\kappa$ over 0.3 when using the generalized conditional probability approach for binary co-occurrence matrices;
    \item don't use country-level or establishment-level co-occurrences, but if available, co-occurrences at intermediate levels of aggregation.
\end{itemize}

\emph{When constructing density variables}
\begin{itemize}
    \item don't use binarized prevalences;
    \item don't use the raw size of an industry in a city, but compare this size against a benchmark; 
    \item don't use highly skewed prevalences;
    \item don't use (negative) unrelatedness, but only focus on industries that are (positively) related to the focal industry;
\end{itemize}

\begin{figure}[!h]
	\begin{center}
	\includegraphics[scale=.90]{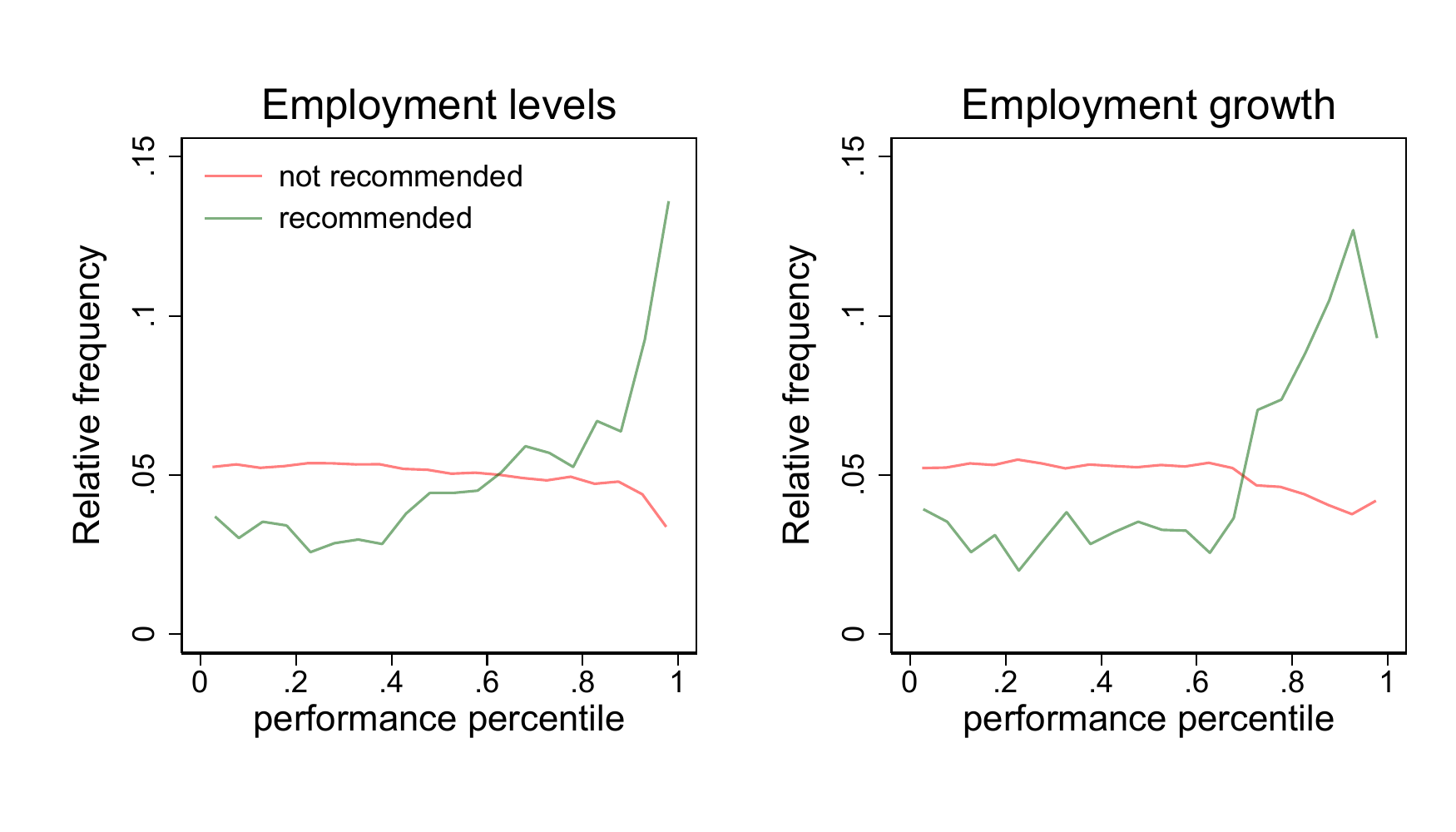}
    \mycaption{Performance histograms.}{Red lines show relative frequencies across performance percentiles of specifications that are expected to perform poorly: i.e, specifications relying on (1) country or establishment portfolios to calculate relatedness (2) raw size or untransformed RCAs to measure industry-unit prevalence, (3) fixed effects or Poisson models to define industry-unit prevalence benchmarks, (4) $\kappa>0.3$ in generalized conditional probability approaches, (5) binarized, raw size or untransformed RCAs in density calculations, (6) fixed effects or Poisson models to define industry-city prevalence benchmarks (7) negative elements of relatedness matrices when calculating density. Green lines show relative frequencies for all other specifications.}\label{fig:poorspecs}   
	\end{center}
\end{figure}

Based on these recommendations, poor  specifications account for about 85\% of the grid. Figure \ref{fig:poorspecs} compares the performance of these poor specifications to the remaining specifications. Red lines show histograms over the performance percentiles of poor specifications, green lines of all remaining specifications. The figure corroborates the recommendations formulated above: the remaining specifications are much more likely to perform well in either prediction task than the ones we expect to perform poorly.

\newpage
\section{Definition of four groups of industries} \label{app:ind_groups}

We divide the 415 3-digit SIC codes into 4 groups that should differ in the extent to which their location patterns will follow local capabilities:

\emph{Resource-based industries} are mainly agriculture, forestry, fishing and mining related activities, which depend on the existence of local resources. This group consists of the following 3-digit industries:

\begin{quote}
011, 013, 016, 017, 018, 019, 021, 024, 025, 027, 029, 071, 072, 074, 075, 076, 078, 081, 083, 085, 091, 092, 097, 101, 102, 103, 104, 106, 108, 109, 122, 123, 124, 131, 132, 138, 141, 142, 144, 145, 147, 148, 149
\end{quote}

\emph{Non-traded services} include construction, transportation, various wholesale and retail trades. This group consists of the following 3-digit industries:

\begin{quote}
152, 153, 154, 161, 162, 171, 172, 173, 174, 175, 176, 177, 178, 179, 401, 411, 412, 413, 414, 415, 417, 421, 422, 423, 431, 441, 442, 443, 444, 448, 449, 451, 452, 458, 461, 472, 473, 474, 478, 491, 492, 493, 494, 495, 496, 497, 521, 523, 525, 526, 527, 531, 533, 539, 541, 542, 543, 544, 545, 546, 549, 551, 552, 553, 554, 555, 556, 557, 559, 561, 562, 563, 564, 565, 566, 569, 571, 572, 573, 581, 591, 592, 593, 594, 596, 598, 599, 641, 651, 653, 654, 655, 721, 722, 723, 724, 725, 726, 729, 751, 752, 753, 754, 762, 763, 764, 769
\end{quote}

\emph{Public-sector industries} are mainly social and governmental activities. This group consists of the following 3-digit industries:

\begin{quote}
801, 802, 804, 805, 808, 809, 821, 823, 829, 832, 833, 835, 836, 839, 861, 862, 863, 864, 865, 866, 869, 881, 899, 911, 912, 913, 919, 921, 922, 931, 941, 943, 944, 945, 951, 953, 961, 962, 963, 964, 965, 966, 971, 972
\end{quote}

All remaining industries were classified as \emph{traded industries}. They mainly consist of manufacturing industries and tradable services. This group consists of the following 3-digit industries:

\begin{quote}
201, 202, 203, 204, 205, 206, 207, 208, 209, 211, 212, 213, 214, 221, 222, 223, 224, 225, 226, 227, 228, 229, 231, 232, 233, 234, 235, 236, 237, 238, 239, 241, 242, 243, 244, 245, 249, 251, 252, 253, 254, 259, 261, 262, 263, 265, 267, 271, 272, 273, 274, 275, 276, 277, 278, 279, 281, 282, 283, 284, 285, 286, 287, 289, 291, 295, 299, 301, 302, 305, 306, 308, 311, 313, 314, 315, 316, 317, 319, 321, 322, 323, 324, 325, 326, 327, 328, 329, 331, 332, 333, 334, 335, 336, 339, 341, 342, 343, 344, 345, 346, 347, 348, 349, 351, 352, 353, 354, 355, 356, 357, 358, 359, 361, 362, 363, 364, 365, 366, 367, 369, 371, 372, 373, 374, 375, 376, 379, 381, 382, 384, 385, 386, 387, 391, 393, 394, 395, 396, 399, 481, 482, 483, 484, 489, 501, 502, 503, 504, 505, 506, 507, 508, 509, 511, 512, 513, 514, 515, 516, 517, 518, 519, 601, 602, 603, 606, 608, 609, 611, 614, 615, 616, 621, 622, 623, 628, 631, 632, 633, 635, 636, 637, 639, 671, 672, 673, 679, 701, 702, 703, 704, 731, 732, 733, 734, 735, 736, 737, 738, 781, 782, 783, 784, 791, 792, 793, 794, 799, 803, 806, 807, 811, 822, 824, 841, 842, 871, 872, 873, 874
\end{quote}

\newpage
\section{Distributing employees across an establishment's industries} \label{app:est_emp_share}

The D\&B database includes up to six different industry codes for each establishment that are listed in decreasing order of importance. However, the database does not offer an estimate of the number of employees associated with each industry. To impute such an estimate, we proceed as follows.

For establishments with only one industry code, we associate the total number of employees in the establishment with that industry code. For establishments with multiple industry codes, we need to estimate a function $r_i=f(N_{industry}, rank_i)$ that maps the number of industries $N_{industry}$ and the rank of an industry code $rank_i$ to a ratio $r_i$, such that $\sum_i r_i=1$. For example, for an establishment with a total number of employees $E$ and three industry codes, the mapping would require that we estimate $E_1=E f(3, 1)$, $E_2=E f(3, 2)$ and $E_3=E f(3, 3)$.

To complete this task, we assume that the distribution of employment across industries at the establishment level is similar to that at the firm level. That is, we estimate the function above from the shares of employees for primary industry codes in US firms. Take $f(3, 1)$ as an example, the ratio is estimated as the normalized geometric mean of the largest employment shares in firms with three different primary industry codes. The results of this estimation are shown in Table \ref{tab:est_emp_share}.

\begin{table}
    \begin{center}
        \begin{tabular}{ccc}
\hline 
 $N_{industry}$ &  $rank_i$ &  $r_i$ \\
\hline 
         1 &    1 &   1.000 \\
         2 &    1 &   0.803 \\
         2 &    2 &   0.197 \\
         3 &    1 &   0.735 \\
         3 &    2 &   0.202 \\
         3 &    3 &   0.063 \\
         4 &    1 &   0.686 \\
         4 &    2 &   0.203 \\
         4 &    3 &   0.082 \\
         4 &    4 &   0.029 \\
         5 &    1 &   0.647 \\
         5 &    2 &   0.205 \\
         5 &    3 &   0.089 \\
         5 &    4 &   0.042 \\
         5 &    5 &   0.017 \\
         6 &    1 &   0.610 \\
         6 &    2 &   0.205 \\
         6 &    3 &   0.098 \\
         6 &    4 &   0.050 \\
         6 &    5 &   0.025 \\
         6 &    6 &   0.012 \\
\hline 
\end{tabular}
    \end{center}
    \mycaption{Estimation of industrial employment shares within establishments level}{}\label{tab:est_emp_share}
\end{table}

\newpage
\section{Comparing relatedness matrices} \label{app:correlation_relmat}

\begin{figure} [t!]
	\begin{center}
	\includegraphics[scale=.6]{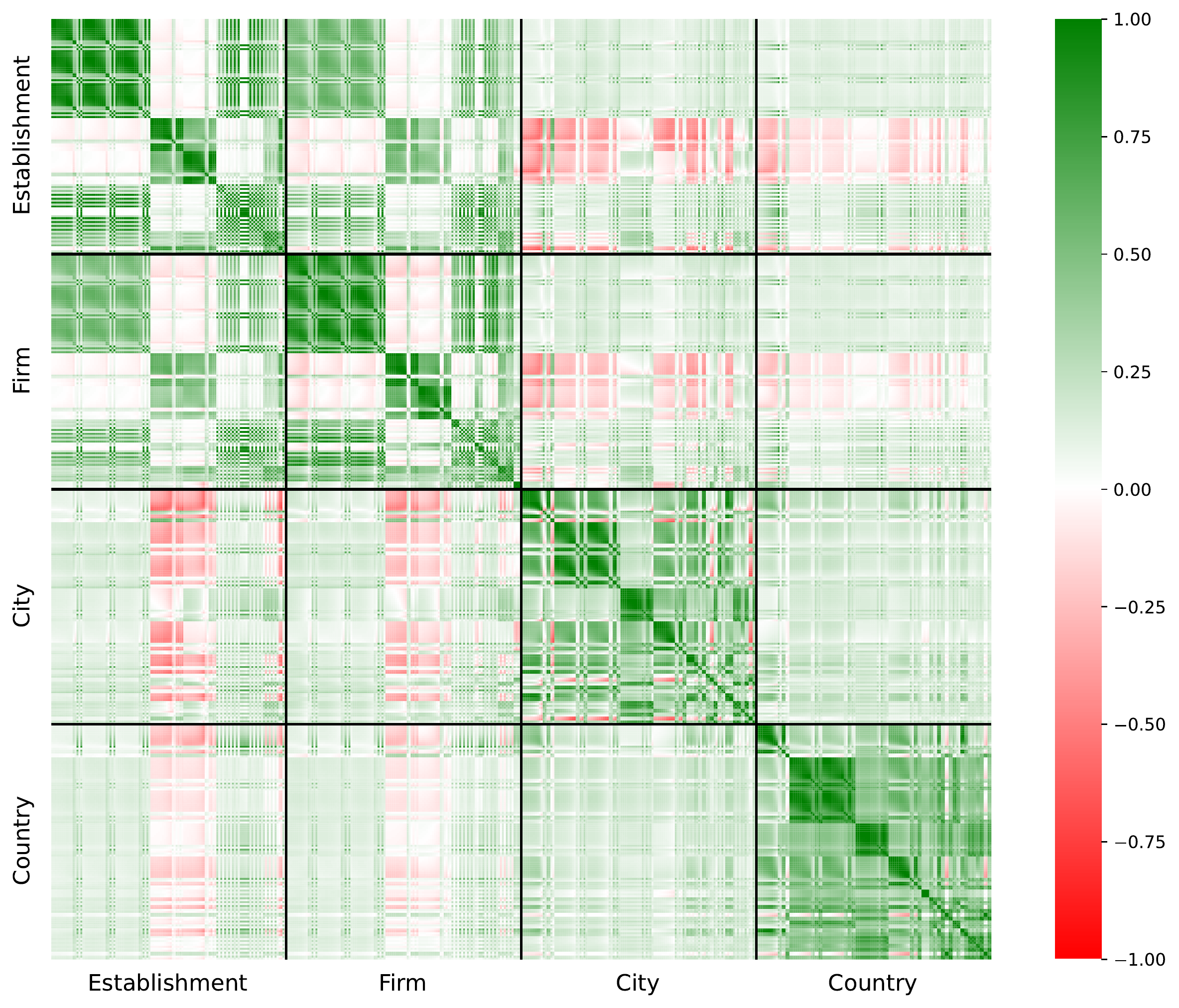}
    \mycaption{Pairwise Pearson correlation of relatedness matrices}{Observation $(i,j)$ represents the correlation between two vectors that stack the columns of the relatedness matrices associated with specification $i$ and $j$. Relatedness matrices are first grouped by the productive entities in which industry combinations are observed. Within these groupings, specifications are collected by the definition of industry prevalence, the metric used to convert coprevalence or co-occurrence information into a relatedness matrix and finally the exact parameter settings used therein.}    \label{fig:corrplot}
	\end{center}
\end{figure}

Figure \ref{fig:corrplot} compares relatedness matrices across specifications by plotting the correlations between all different relatedness matrices that were created in the specification grid. Specifications are ordered along both axes first by the productive unit in which relatedness was measured and then by other aspects of the specification. 

The large diagonal blocks show that different productive units yield different relatedness matrices, suggesting that they identify different types of economies of scope. Within a productive unit, there are three pronounced smaller blocks. The first two correspond to binarized specifications, the third to continuous specifications. Moreover, the block in the middle is quite distinct from the other two blocks. This block contains binarized specifications, based on residuals from first fixed effects and then a simple OLS regression. 

The exact details of the specification matter less and less as we move from establishments to higher-order productive units. At the country level, relatedness matrices become very similar regardless of the specification used.

\newpage
\section{Additional results on density}\label{app:denspairs}

\begin{table}[h!]
    \begin{center}
        \begin{tabular}{l*{7}{c}}
\hline\hline
  \textbf{productive unit}          &\multicolumn{1}{c}{(1)}&\multicolumn{1}{c}{(2)}&\multicolumn{1}{c}{(3)}&\multicolumn{1}{c}{(4)}&\multicolumn{1}{c}{(5)}&\multicolumn{1}{c}{(6)}\\
\hline
\multicolumn{7}{c}{\textit{Effect of density on employment levels}}\\

\hline  
establishment &       0.116         &       0.114\sym{**} &                     &       0.452\sym{***}&                     &                     \\
            &    (0.0606)         &    (0.0440)         &                     &    (0.0664)         &                     &                     \\
firm &       0.630\sym{***}&                     &       0.207\sym{***}&                     &       0.592\sym{***}&                     \\
            &    (0.0538)         &                     &    (0.0429)         &                     &    (0.0573)         &                     \\
city &                     &       0.799\sym{***}&       0.733\sym{***}&                     &                     &       0.795\sym{***}\\
            &                     &    (0.0260)         &    (0.0284)         &                     &                     &    (0.0342)         \\
country &                     &                     &                     &       0.372\sym{***}&       0.267\sym{***}&       0.129\sym{***}\\
            &                     &                     &                     &    (0.0218)         &    (0.0229)         &    (0.0136)         \\
\hline\(N\)    &      384,705         &      384,705         &      384,705         &      384,705         &      384,705         &      384,705           \\
\(R^{2}\)   &       0.739         &       0.786         &       0.788         &       0.733         &       0.747         &       0.787         \\
\hline 
\multicolumn{7}{c}{\textit{Effect of density on employment growth}} \\ \hline 
establishment &     0.00220\sym{*}  &     0.00378\sym{*}  &                     &     0.00448\sym{*}  &                     &                     \\
            &  (0.000947)         &   (0.00174)         &                     &   (0.00180)         &                     &                     \\
[1em]
firm &     0.00777\sym{***}&                     &     0.00728\sym{***}&                     &     0.00773\sym{***}&                     \\
            &   (0.00137)         &                     &   (0.00147)         &                     &   (0.00135)         &                     \\
city &                     &     0.00583\sym{**} &     0.00428\sym{*}  &                     &                     &     0.00573\sym{**} \\
            &                     &   (0.00188)         &   (0.00192)         &                     &                     &   (0.00193)         \\
country &                     &                     &                     &     0.00450\sym{***}&     0.00297\sym{**} &     0.00358\sym{**} \\
            &                     &                     &                     &   (0.00106)         &   (0.00108)         &   (0.00125)         \\

\hline 
\(N\)         &      245,395         &      245,395         &      245,395         &      245,395         &      245,395         &      245,395          \\
\(R^{2}\)   &       0.068         &       0.066         &       0.069         &       0.066         &       0.069         &       0.066         \\
\hline\hline
\end{tabular}
    \end{center}
  \mycaption{Redundancy and complementarity in pairs of density variables (preferred continuous specification)}{Standard errors in parentheses, \sym{*}: \(p<0.05\), \sym{**}: \(p<0.01\), \sym{***}: \(p<0.001\). Upper panel: regression analysis of employment levels, controlling for industry and city size. Lower panel: regression analysis of employment growth, controlling for mean reversion effects, industry and city size. The coefficients reflect the effects of density measured using the productive units listed in the first column. All density measures are based on our preferred continuous specification.}\label{tab:densitypairs_con}
\end{table}

\begin{table}[h!]
    \begin{center}
        \begin{tabular}{l*{5}{c}}
\hline\hline
            &\multicolumn{1}{c}{Establishment}&\multicolumn{1}{c}{Firm}&\multicolumn{1}{c}{City}&\multicolumn{1}{c}{Country}\\
\hline
\multicolumn{5}{c}{\textit{Employment levels}} \\
\hline
$\text{density}_{mass}$    &     0.663***   &     0.829***   &     1.154***   &     0.338*** \\
               &   (0.128)      &  (0.0619)      &  (0.0531)      &  (0.0498)    \\
$\text{density}_{var}$     &    -0.108      &    -0.161*     &    -0.334***   &     0.194*   \\
               &   (0.155)      &  (0.0698)      &  (0.0784)      &  (0.0756)    \\
\hline
  \(N\)                &      384705         &      384705         &      384705         &      384705             &    \\
  \(R^2\)                &     0.715     &      0.740     &      0.788   &        0.710              &    \\
\hline
\multicolumn{5}{c}{\textit{Employment growth}}\\
\hline
$\text{density}_{mass}$   &    0.00376     &    0.00672***   &    0.0112***  &    0.0130***  \\
                          &  (0.00207)     &  (0.00169)      & (0.00282)     & (0.00170)     \\
$\text{density}_{mass}$   &    0.00238     &    0.00319      &  -0.00442     &  -0.00875***  \\
                          &  (0.00201)     &  (0.00208)      & (0.00355)     & (0.00261)     \\
\hline
\(N\)                &      245395         &      245395             &      245395         &      245395       \\
 \(R^2\)                &     0.064      &     0.069        &   0.065     &      0.065   \\
\hline\hline
\end{tabular}                    
    \end{center}
  \mycaption{Related variety versus mass of related activity (preferred continuous specification).}{Standard errors in parentheses, \sym{*}: \(p<0.05\), \sym{**}: \(p<0.01\), \sym{***}: \(p<0.001\). Upper panel: regression analysis of employment levels, controlling for industry and city size effects. Lower panel: regression analysis of employment growth, controlling for mean reversion, industry and city size effects. The coefficients reflect the effects of density based on relatedness measured in the productive units listed in the columns, where $\text{density}_{mass}$ uses continuous information on an industry's prevalence in a city, and $\text{density}_{var}$ continuous information on whether or not an industry is significantly present in the city.}\label{tab:variety_vs_mass_con}
\end{table}

\newpage
\section{Grid search figures}\label{APP:gridfigs}

In this section, we plot distributions of performance statistics for different choices of parameter $\kappa$ in the conditional probability metric and for different numbers of neighbors in the construction of the density measure.




\begin{figure}[h!] 
	\begin{center}
	\includegraphics[scale=1.1]{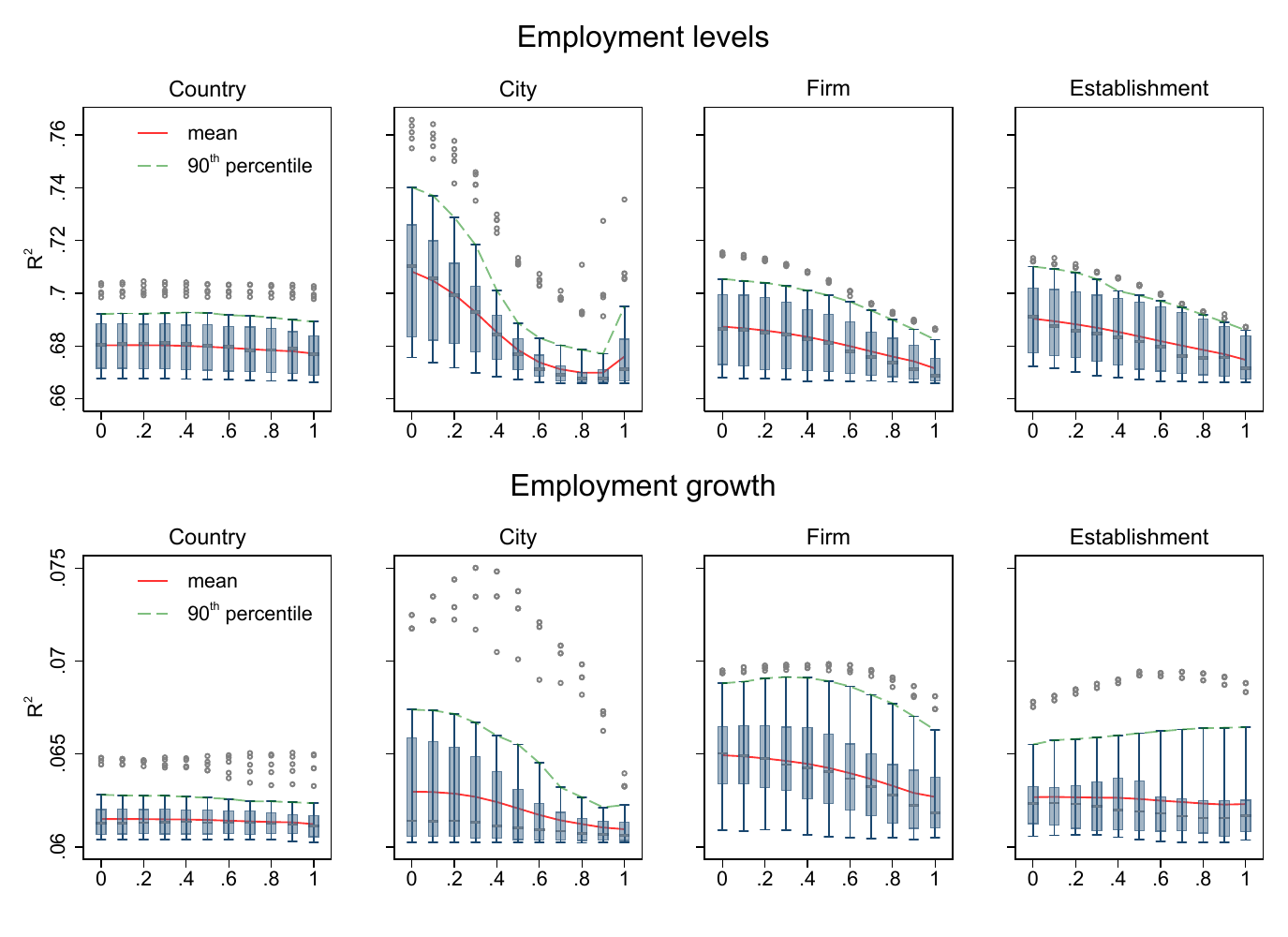}
    \mycaption{Conditional probability.}{The graphs show the out-of-sample $R^2$ for specification that rely on eq. (\ref{eq:generalized_min_prob}) to estimated relatedness for different values of $\kappa$, where $\kappa=0$ corresponds to the minimum conditional probability approach in \cite{hidalgo2007product}. The boxes display the interquartile range, the whiskers extend from the \nth{10} to the \nth{90} percentile. The gray circles mark the $R^2$ of the five best specifications.}    \label{fig:cond_prob}
	\end{center}
\end{figure}


\begin{figure}[t!]
	\begin{center}
	\includegraphics[scale=1.1]{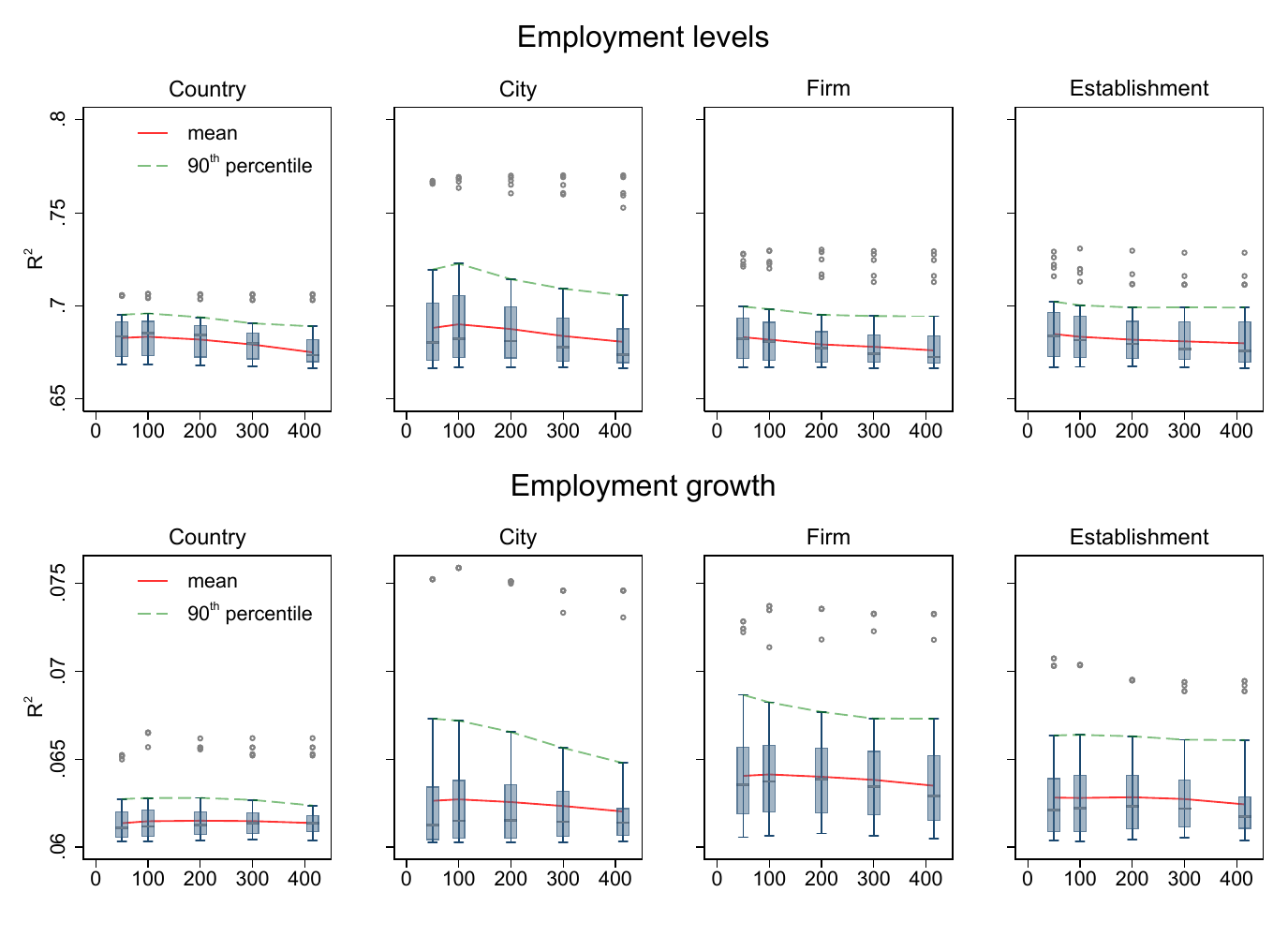}
    \mycaption{Number of neighbors.}{The graphs show the out-of-sample $R^2$ for different numbers of neighbors considered in the calculation of density, $d_{irt}$. The boxes display the interquartile range, the whiskers extend from the \nth{10} to the \nth{90} percentile. The gray circles mark the $R^2$ of the five best specifications.}   \label{fig:num_neighbors}
	\end{center}
\end{figure}

\newpage
\section{Industry spaces}\label{APP:industry_spaces}

In this appendix, we visualize industry spaces for each productive unit type using relatedness matrices from our preferred binary and continuous specifications. To do so, we follow the approach of \cite{hidalgo2007product} to create a network visualization from the dense relatedness matrix. That is, we first calculate the maximum spanning tree to extract the backbone of the relatedness matrix, and then add links with high relatedness  until the number of edges equals three times the number of nodes. Nodes in the networks represent industries, ties strong relatedness connections. Nodes are furthermore colored by the high level sectors to which their industries belong.

\begin{figure} 
	\begin{center}
	\includegraphics[scale=.55]{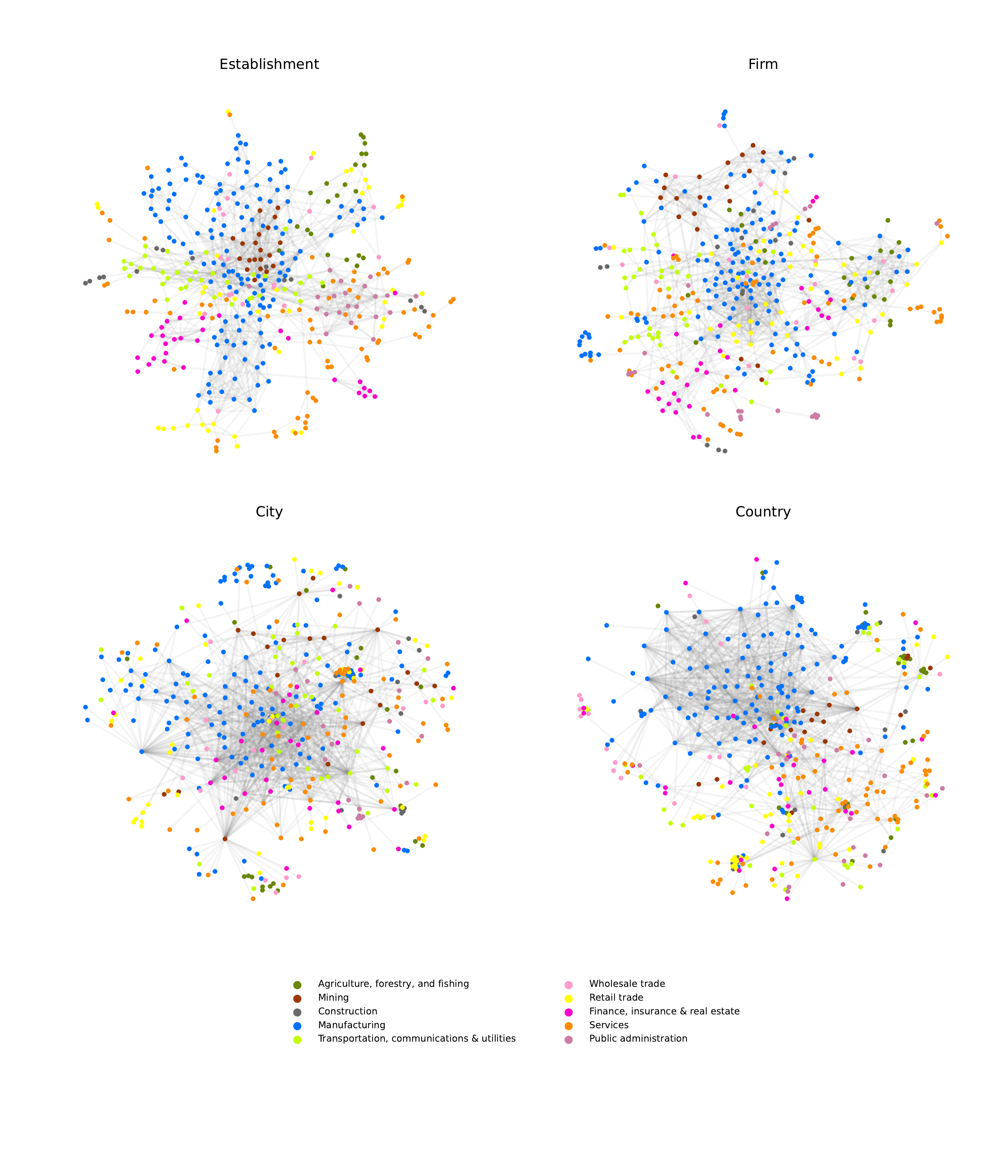}
    \caption{Industry spaces - binary.}\label{fig:indspace_binary}
	\end{center}
\end{figure}

\begin{figure}
	\begin{center}
	\includegraphics[scale=.55]{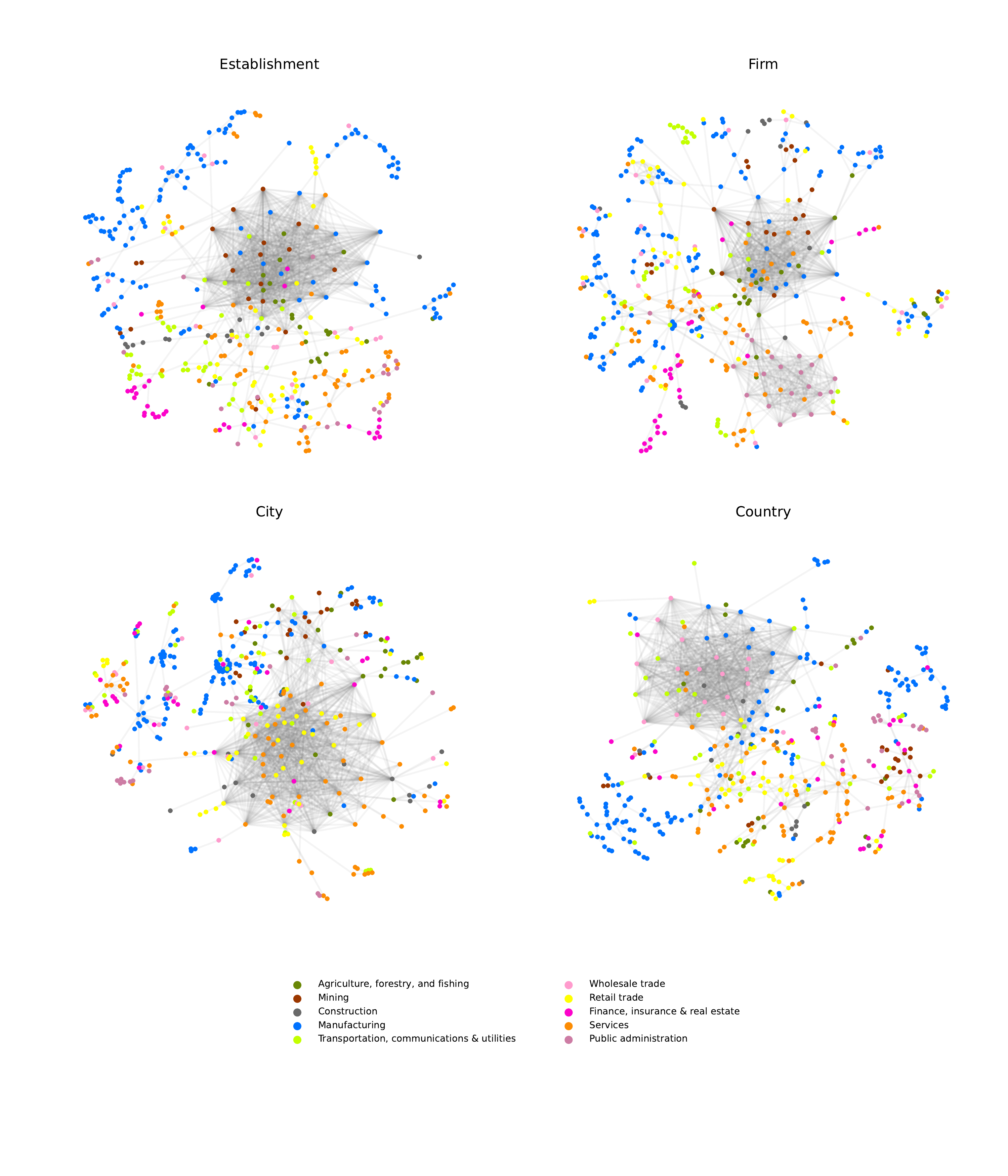}
    \caption{Industry spaces - continuous.}\label{fig:indspace_continuous}
	\end{center}
\end{figure}

\end{document}